\pdfoutput=1
\documentclass[journal=jpcbfk,manuscript=article]{achemso}
\setkeys{acs}{articletitle=true}

\usepackage[version=3]{mhchem} % Formula subscripts using \ce{}

\usepackage{mathtools,pdfpages,graphicx,pdflscape}
\usepackage{lscape}

\newcommand*{\citen}[1]{%
\begingroup
\romannumeral-`\x % remove space at the beginning of \setcitestyle
\setcitestyle{numbers}%
\cite{#1}%
\endgroup   
}

\author{Adrien Lerbret}
\affiliation[Univ. Bourgogne Franche-Comt\'e]
{Univ. Bourgogne Franche-Comt\'e, AgroSup Dijon, UMR A 02.102 - PAM - Proc\'ed\'es Alimentaires et Microbiologiques, F-21000 Dijon, France}
\email{adrien.lerbret@u-bourgogne.fr}
\phone{+33 (0)3-80-77-40-61}
\author{Fr\'ed\'eric Affouard}
\affiliation[Univ. Lille]
{Univ. Lille, CNRS, UMR 8207 - UMET - Unit\'e Mat\'eriaux Et Transformations, F-59000 Lille, France}

\title[\texttt{achemso} demonstration]
{Molecular Packing, Hydrogen Bonding, and Fast Dynamics in Lysozyme/Trehalose/Glycerol and 
Trehalose/Glycerol Glasses at Low Hydration}

\begin{document}
\begin{singlespacing}
%%%%%%%%%%%%%%%%%%%%%%%%%%%%%%%%%%%%%%%%%%%%%%%%%%%%%%%%%%%%%%%%%%%%%
%% The manuscript does not need to include \maketitle, which is
%% executed automatically.  The document should begin with an
%% abstract, if appropriate.  If one is given and should not be, the
%% contents will be gobbled.
%%%%%%%%%%%%%%%%%%%%%%%%%%%%%%%%%%%%%%%%%%%%%%%%%%%%%%%%%%%%%%%%%%%%%
\begin{abstract}
Water and glycerol are well-known to facilitate the structural relaxation of 
amorphous protein matrices. However, several studies evidenced that they may also limit fast 
($\sim$ pico-nanosecond, ps-ns) and small-amplitude ($\sim$ \AA \/) motions of proteins, which govern 
their stability in freeze-dried sugar mixtures. To determine how they interact with proteins
and sugars in glassy matrices and, thereby, modulate their fast dynamics, we performed molecular dynamics (MD) 
simulations of lysozyme/trehalose/glycerol (LTG) and trehalose/glycerol (TG) mixtures at low glycerol and water 
concentrations. Upon addition of glycerol and/or water, the glass transition temperature, T$_{\textrm{g}}$, 
of LTG and TG mixtures decreases, the molecular packing of glasses is improved, and the mean-square displacements (MSDs) of 
lysozyme and trehalose either decrease or increase, depending on the time scale and on the temperature considered. 
A detailed analysis of the hydrogen bonds (HBs) formed between species reveals that water and glycerol may 
\textit{antiplasticize} the fast dynamics of lysozyme and trehalose by increasing the total number 
and/or the strength of the HBs they form in glassy matrices.
\end{abstract}

\begin{figure}[htbp]
\includegraphics[width=14cm,clip=true]{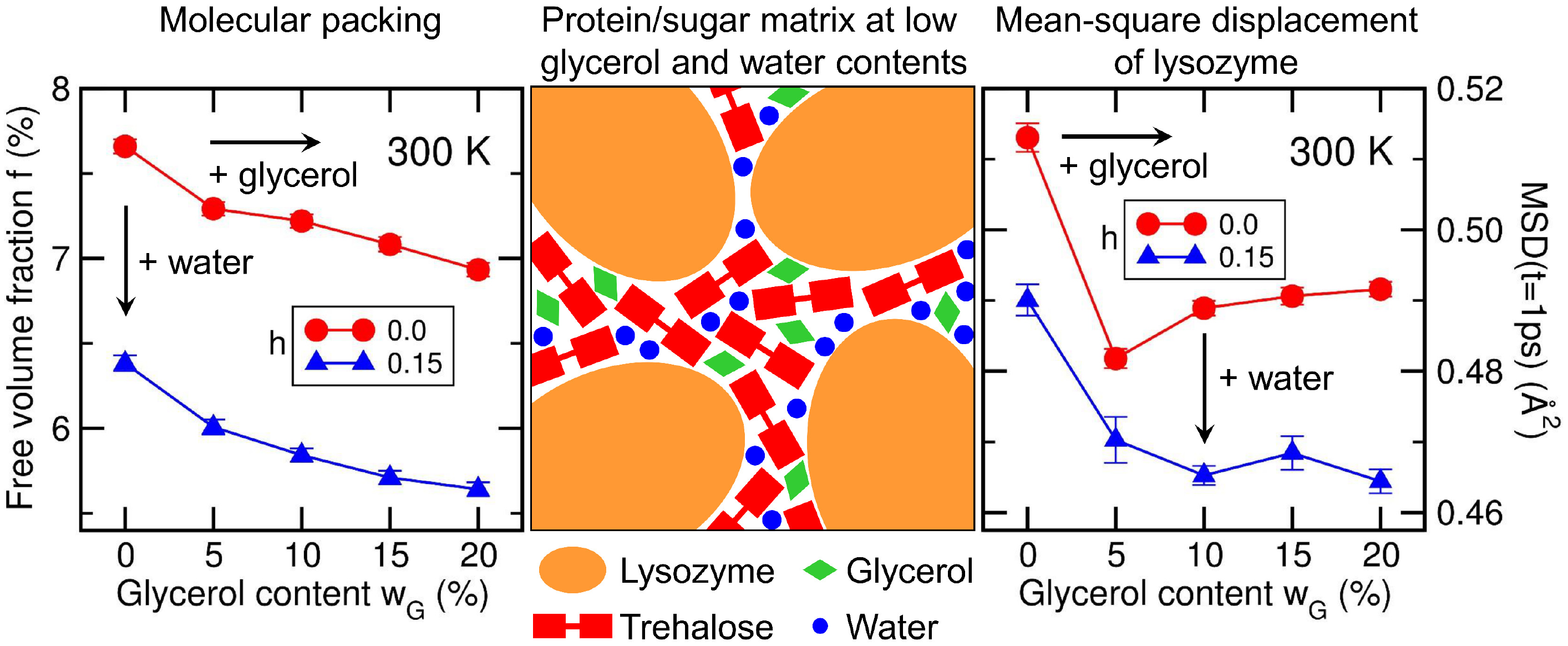}
\end{figure}

\newpage

\section{Introduction}

Most globular proteins are very sensitive to the various stresses that they might 
experience upon purification, processing and long-term storage (changes in temperature, pressure, pH, hydration level, or
ionic strength)~\cite{Wang1999,Wang2000,Arakawa2001,Manning2010,Ohtake2011}.
Adding solutes such as sugars, polyols, polymers, amino acids or salts is a common practice to improve their stability, 
both in the liquid and in the solid states~\cite{Wang1999,Wang2000,Manning2010,Ohtake2011,Soltanizadeh2014,Cicerone2015}. 
However, the molecular mechanisms by which these compounds preserve proteins from degradation have not been fully deciphered, 
in spite of intensive research. In the case of sugars and polyols, many studies proposed that 
proteins are stabilized thermodynamically by the \textit{preferential} exclusion of these \textit{co-solutes} 
from the protein/solvent interface in aqueous solution~\cite{Gekko1981,Timasheff2002}, while at low water contents 
proteins would be preserved both by the formation of numerous protein-solute hydrogen bonds (HBs)~\cite{Carpenter1989,Crowe1998,Allison1999,Sakurai2008} 
and by the vitrification of the mixture~\cite{Green1989}. Yet, there are still many open questions, 
in particular on how water, polyols, and sugars interact with each other and/or with proteins in complex glassy mixtures.\\
Low-molecular-weight compounds such as water, glycerol (C$_{\textrm{3}}$H$_{\textrm{8}}$O$_{\textrm{3}}$), or 
sorbitol (C$_{\textrm{6}}$H$_{\textrm{14}}$O$_{\textrm{6}}$), are often regarded as \textit{plasticizers} of 
carbohydrate and protein 
matrices~\cite{Roos1991b,Hancock1994,Lourdin1998,Paciaroni2002,Verbeek2010,Townrow2010,Roussenova2010,Bellavia2011,Roussenova2014,Roussenova2014b,Weng2015,Ubbink2016}, 
since the addition of these small molecules (sometimes referred to as \textit{diluents}~\cite{Lourdin1998,Seow1999,Cicerone2004,Townrow2010,Roussenova2010,Cicerone2012,Roussenova2014,Roussenova2014b,Ubbink2016}) usually decreases their glass transition temperature, T$_{\textrm{g}}$, as well as 
their elastic moduli, and increases the free volume and the water and oxygen permeabilities. 
However, several studies have demonstrated that they can also act as \textit{antiplasticizers}, especially at low temperatures and 
concentrations~\cite{Lourdin1997,Seow1999,Cicerone2004,Chang2005_2,Townrow2010,Roussenova2010,Cicerone2012,Roussenova2014,Roussenova2014b,Ubbink2016}.
Ubbink and coworkers thoroughly investigated structural and thermodynamics aspects of antiplasticization on various glassy carbohydrate matrices
(maltopolymer-maltose blends, maltodextrin, etc.) and showed that the antiplasticization induced by the addition of water 
and/or glycerol stems from the decrease of the average hole volume, $v_{\textrm{h}}$, 
as probed by positron annihilation lifetime spectroscopy (PALS), as well as from the decrease of the specific volume, V$_{\textrm{sp}}$, 
estimated using gas pycnometry (see ref.~[\citen{Ubbink2016}] and references therein). 
Moreover, their results revealed that the antiplasticization of glassy carbohydrate and biopolymer matrices 
by low-molecular-weight diluents manifests as a strengthening of H-bonding interactions. By means of Fourier-transform infrared (FTIR) 
spectroscopy, they indeed observed a low-frequency shift of the O-H stretching vibration frequency, $\nu_{OH}$, upon addition of 
low amounts of water and/or glycerol~\cite{Roussenova2014,Ubbink2016}, and found a linear correlation between changes 
in $\nu_{OH}$ and changes in $v_{\textrm{h}}$, thereby evidencing the interdependence between molecular interactions and molecular 
packing in these H-bonding systems~\cite{Ubbink2016}.
Such modifications of structural properties and of H-bonding interactions probably account for the changes
in the fast dynamics of glassy carbohydrate matrices described in the literature 
using various techniques. Cicerone and Soles evidenced by means of neutron scattering experiments that the addition of 5 wt. \% of glycerol 
strongly reduces the mean-square displacement (MSD), $<u^{\textrm{2}}>$, of fast ($>$ 200 MHz) motions of 
trehalose molecules (C$_{\textrm{12}}$H$_{\textrm{22}}$O$_{\textrm{11}}$) over a broad temperature range 
($\sim$ 100-350 K)~\cite{Cicerone2004}. They derived an effective 
spring constant much larger for this mixture than for pure freeze-dried trehalose and, interestingly, found an inverse 
relationship between the $<u^{\textrm{2}}>$ of glassy protein/sugar matrices and the degradation rates of the embedded
enzymes (horseradish peroxidase and yeast alcohol dehydrogenase)~\cite{Cicerone2004}. More recently, Cicerone and Douglas 
confirmed such a relationship for a large series of proteins mixed with either trehalose or sucrose, and suggested that 
the degradation of proteins in carbohydrate glasses is governed by small amplitude, fast motions ($\beta$-relaxation 
processes) of the glassy matrix rather than by structural, $\alpha$ relaxation~\cite{Cicerone2012}, thus 
showing that the antiplasticization of fast motions at sub-T$_{\textrm{g}}$ temperatures is particularly relevant for 
biostabilization purposes~\cite{Cicerone2012,Ubbink2016}. 
The pioneering results of Cicerone and Soles~\cite{Cicerone2004} have triggered many studies trying to further
understand the antiplasticizing effect of glycerol on 
trehalose~\cite{Dirama2005,Anopchenko2006,Riggleman2008,Magazu2010,Obrzut2010,Averett2012,Weng2015}, 
as well as the enhanced stability that 
such mixtures may confer to embedded proteins~\cite{Curtis2006,Cicerone2012,Barreca2013,Bellavia2014}.

The antiplasticization of trehalose motions by glycerol has been observed at various concentrations 
and temperatures in experimental and numerical studies. 
From the secondary relaxation times, $\tau$, obtained by means of dielectric relaxation spectroscopy 
measurements over broad ranges of temperature and glycerol concentration, Anopchenko et al. determined the 
magnitude of antiplasticization, defined as the ratio $\tau_{TG}$/$\tau_{T}$ ($\beta$-relaxation time in the 
trehalose/glycerol, TG, mixture over that in pure trehalose, T), and found maxima in the 20-30 wt \% concentration 
range~\cite{Anopchenko2006,Obrzut2010}. Similarly, the molecular dynamics (MD) simulation results of 
Averett et al. showed that the $<u^{\textrm{2}}>$ of TG mixtures is minimum between 20 and 50 wt.~\%~\cite{Averett2012}.
In contrast, Dirama~{et al.} and Magazu~{et al.} observed minima of $<u^{\textrm{2}}>$ at a glycerol weight concentration 
of 5 and 2.5 \% at 300~K and 100~K, respectively. Furthermore, Weng and Elliott found by means of 
dynamical mechanical analysis (DMA) that the fragility index, m, of dehydrated TG mixtures decreases 
in a non-monotonic manner in the 0-20 wt.~\% range, with a local minimum between 5 and 12.5~wt.~\%~\cite{Weng2015}.

Discrepancies also emerged from studies on proteins in TG mixtures:
the $<u^{\textrm{2}}>$ of RNase was minimum at a glycerol concentration of 12.2 wt. \%. in the MD study of Curtis et al., 
whereas Bellavia et al. evidenced from Raman spectroscopy that 
the denaturation temperature, T${_\textrm{m}}$, of lysozyme exhibits a slight
maximum at a glycerol concentration of 5~\%~\cite{Bellavia2014}.
The discrepancies among all these studies probably arise from the diversity of the techniques employed 
(neutron scattering, Raman and dielectric relaxation spectroscopies, DMA, MD simulation) to investigate TG mixtures 
or proteins in TG matrices, as well as from the various water contents, proteins or force fields considered in experiments and simulations.
For instance, the differences that emerge among the MD
studies on TG mixtures~\cite{Dirama2005,Riggleman2008,Magazu2010,Averett2012} indicate a strong dependence of results
on the force fields used to simulate trehalose and glycerol, which might not represent accurately enough pure trehalose and glycerol, 
and/or trehalose/glycerol mixtures. 
This is actually why Averett et al. derived specifically a force field for glycerol that reproduces satisfactorily
the experimental temperature dependence of bulk glycerol's density~\cite{Averett2012}. 
Using this revised force field for glycerol, they did not observe any slight density maximum upon addition of glycerol to trehalose, 
in contrast to previous MD studies by Dirama et al.~\cite{Dirama2005} and by Magazu et al.~\cite{Magazu2010}, at glycerol
contents of 5 wt. \% and of 2.5 wt. \%, respectively. Given the discrepancies among MD studies, and because no experimental data 
exists on the molecular hole volume or on the specific volume (that is, on density) for trehalose/glycerol glasses, it is not possible 
to definitely ascribe the antiplasticizing effect of glycerol on trehalose fast motions to an improved molecular packing, 
as found for other carbohydrate matrices by Ubbink and coworkers~\cite{Ubbink2016}. However, it probably involves 
to some extent changes in the HB network of trehalose. Dirama et al. actually observed that the occupancy of HBs is maximum at a
glycerol content of 5 wt. \% and suggested that the minimized amplitude of trehalose 
motions induced by glycerol originates from the simultaneous formation of multiple T-G HBs, with glycerol molecules being sometimes 
able to \textit{bridge} the rings of trehalose~\cite{Dirama2005}. Curtis et al. also observed that the maximum effect 
of glycerol correlates with a maximum number of T-G HBs~\cite{Curtis2006}. These simulation results imply that glycerol induces
a strengthening of H-bonding interactions, in line with the FTIR study of Roussenova et al. on maltooligomer matrices~\cite{Roussenova2014}.

To further understand how glycerol and water interact with trehalose and proteins and modulate their fast dynamics 
in glassy mixtures, we performed MD simulations of lysozyme/trehalose/glycerol (LTG) and trehalose/glycerol (TG) matrices at various 
glycerol concentrations (w$_{\textrm{G}}$ = 0, 5, 10, 15, and 20 wt. \% of the total mass of excipients) and water contents 
($h$ = 0.0, 0.075, and 0.15 in g of water per g of protein or excipient).
The comparison of LTG and TG matrices allows us to determine whether the improved protein stability in presence of glycerol
observed experimentally~\cite{Cicerone2004,Cicerone2012} simply results from the antiplasticization of trehalose motions, 
assuming that the protein dynamics follows that of the sugar matrix~\cite{Caliskan2004,Cornicchi2006}, or whether it also stems 
from specific protein-glycerol interactions not present in trehalose/glycerol binary mixtures. 
Moreover, it is necessary to consider the matrices at several 
low water contents, since lyophilized powders usually contain a substantial amount of residual water, which obviously 
modulates the concentration of glycerol at which maximal antiplasticization is observed~\cite{Anopchenko2006}. 
Water is indeed known to strongly modify the properties of glassy matrices, even at 
low concentration~\cite{Roos1991b,Hancock1994,Seow1999,Verbeek2010}, and it may also play a role in the antiplasticization 
of carbohydrate matrices~\cite{Lourdin1997,Townrow2010,Roussenova2010,Roussenova2014,Roussenova2014b,Ubbink2016}. 
In a previous comparative study of lysozyme (L), trehalose (T), and lysozyme/trehalose (LT) glassy matrices at 
300 K~\cite{Lerbret2012}, we actually evidenced that water at $h$ = 0.075 and 0.15 may have an antiplasticizing effect on 
lysozyme and trehalose motions at the pico-nanosecond (ps-ns) time scale. In the present work, we investigate LTG and TG matrices 
at both 300~K and 100~K, given that antiplasticization is known to depend on temperature~\cite{Anopchenko2006,Obrzut2010}. 
We first estimate the glass transition temperature, T$_{\textrm{g}}$, of the simulated systems, next compare their molecular 
packing, then determine the mean-square displacements (MSDs) of lysozyme and trehalose, and finally characterize intermolecular 
interactions through a detailed analysis of the intermolecular HBs formed between species.

\newpage

\section{Simulation details}

LTG and TG glasses were prepared from the starting structures of the lysozyme/trehalose (LT) and trehalose (T) anhydrous matrices 
from our previous study~\cite{Lerbret2012}. Briefly, randomly selected trehalose molecules were first deleted, and glycerol 
and water molecules were then inserted with random positions and orientations. Each system was minimized, heated up to 700~K 
or 650~K, equilibrated at this temperature, then cooled down to 300~K or 100~K, and further equilibrated. 
Production simulations were performed at the equilibrated density in the canonical (N,V,T) ensemble for 25~ns and 5~ns 
at 300~K and at 100~K, respectively. All simulations were performed using the CHARMM program~\cite{Brooks2009}, version c35b1.
Lysozyme molecules were represented using the all-atom 
CHARMM22 force field~\cite{Mackerell1998}, with the CMAP correction for backbone dihedral angles~\cite{Mackerell2004}. 
Trehalose and glycerol molecules were modeled with the CHARMM36 carbohydrate force field~\cite{Guvench2008,Guvench2009,Hatcher2009} 
and the rigid SPC/E model~\cite{Berendsen1987} was considered for water. The covalent bonds involving an hydrogen atom 
and the geometry of water molecules were constrained using the SHAKE algorithm~\cite{Ryckaert1977}. The equation of motions 
was integrated with a timestep of 1~fs. The Langevin piston method~\cite{Nose1983,Hoover1985} was employed to control temperature 
and pressure during simulations. van der Waals interactions were smoothly force-switched~\cite{Steinbach1994} to zero 
between 8 and 10 \AA \/, and Lorentz-Berthelot mixing-rules have been employed for cross-interaction terms. 
The particle mesh Ewald (PME) method~\cite{Essmann1995} has been used to compute electrostatic interactions. 
Full details on simulation parameters and on the preparation of LTG and TG glasses are provided in the Supporting Information.

\newpage

\section{Results and discussion}

\subsection{Glass transition temperature}

\begin{figure}[htbp]
\includegraphics[width=5cm,clip=true]{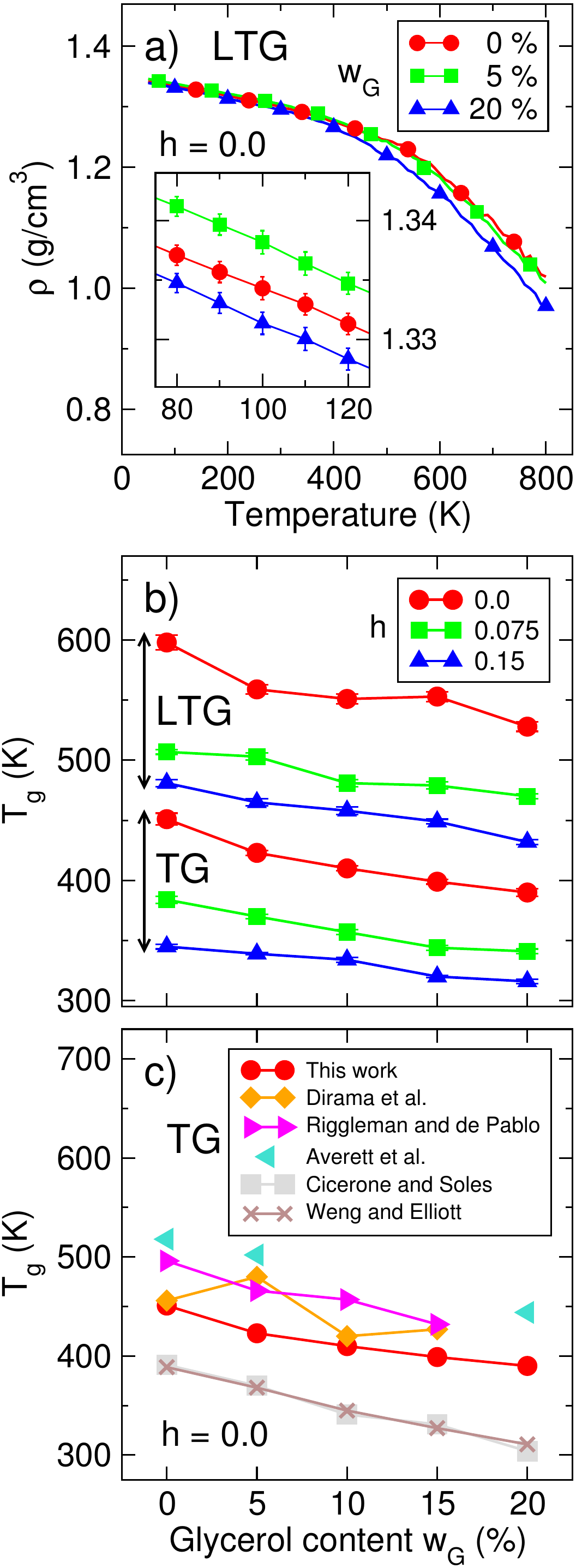}
\caption{\label{density_glass_transition}
(a) Temperature dependence of the density, $\rho$, of anhydrous LTG mixtures at w$_{\textrm{G}}$~=~0, 5, and 20 ~\% upon 
heating from 50 to 800 K at a rate of 0.05 K/ps. The inset shows a zoom on the [75-125~K] temperature range to evidence
the higher $\rho$ of the anhydrous LTG matrix with w$_{\textrm{G}}$ = 5~\% at temperatures below $\sim$ 400~K. 
(b) Glass transition temperature, T$_{\textrm{g}}$, of LTG and TG systems 
as a function of the glycerol content, w$_{\textrm{G}}$, and for the different hydration levels $h$. 
(c) Comparison of the T$_{\textrm{g}}$ for the anhydrous TG mixtures obtained in this work with those determined 
in previous MD simulation~\cite{Dirama2005,Riggleman2008,Averett2012} and experimental~\cite{Cicerone2004,Weng2015} studies.}
\end{figure}

Water and glycerol are well-known to decrease the glass transition temperature, T$_{\textrm{g}}$, of amorphous carbohydrate 
and protein matrices~\cite{Roos1991b,Hancock1994,Lourdin1998,Chen2000,Verbeek2010,Townrow2010,Roussenova2010,Bellavia2011,Roussenova2014,Roussenova2014b,Weng2015}.
As a means to estimate the T$_{\textrm{g}}$ of the simulated LTG and TG systems, we determined the temperature dependence of their
density, $\rho$, upon heating at a rate of 0.05~K/ps (see Figure~\ref{density_glass_transition}a and 
Figure S1 in the Supporting Information), which we then fitted in a similar way as in Averett~\textit{et al.}~\cite{Averett2012} 
(see details and examples of fits in Figure~S2 in the Supporting Information). 
The T$_{\textrm{g}}$ of LTG and TG systems determined in this way decreases significantly 
when the amount of water and/or glycerol increases (Figure~\ref{density_glass_transition}b), in fair agreement with 
simulation~\cite{Dirama2005,Riggleman2008,Averett2012} and experimental~\cite{Chen2000,Cicerone2004,Weng2015} data from 
literature (Figure~\ref{density_glass_transition}c). 
For example, the T$_{\textrm{g}}$ of anhydrous TG mixtures decrease from 451~$\pm$~5~K to 390~$\pm$~3~K in the present simulations and 
from 388.8~$\pm$~0.6~K to 310.8~$\pm$~1.0~K in the DMA study of Weng and Elliott~\cite{Weng2015} when w$_{\textrm{G}}$ increases 
from 0 to 20~\% (Figure~\ref{density_glass_transition}c). To our knowledge, no such data exist for LTG sytems. 
However, Padilla and Pikal determined the T$_{\textrm{g}}$ of the 1:1 L/T freeze dried mixture (LTG matrix with 
w$_{\textrm{G}}$ = 0~\%) to be 404.9~K from Modulated DSC measurements~\cite{Padilla2011}.
Furthermore, we can estimate the T$_{\textrm{g}}$ of 1:1 L/T mixtures to decrease from 363~K to 287~K when $h$ raises from 
0.0 to 0.15, using the Gordon-Taylor parameters obtained by Bellavia~\textit{et al.}~\cite{Bellavia2011}. 
By comparison, 
the T$_{\textrm{g}}$ of the corresponding simulated mixture decreases from 598~$\pm$~6~K to 481~$\pm$~3~K (Figure~\ref{density_glass_transition}b). 
Clearly, the T$_{\textrm{g}}$ of 
simulated LTG and TG mixtures exceed those determined experimentally~\cite{Cicerone2004,Padilla2011,Bellavia2011,Weng2015}. 
Such differences arise in part from the extremely high heating or cooling rates used in MD simulations 
(0.05 K/ps in this study, that is, 3.10$^{\textrm{12}}$ K/min) in comparison with those used experimentally 
(typically $\sim$ 10~K/min)~\cite{Barrat2010,Dirama2005,Riggleman2008,Averett2012}. 
They also stem from the various simulation protocols and force fields employed in MD simulations, 
which, for instance, lead to different densities (see next section). 
Besides, at a given content, the presence of water induces a larger decrease of the T$_{\textrm{g}}$ of LTG and TG systems 
than that of glycerol (Figure~\ref{density_glass_transition}b), in line with experimental data~\cite{Chen2000,Cicerone2004,Weng2015}, thereby 
suggesting that water interacts more strongly with lysozyme and trehalose than glycerol does. 
It must also be pointed out that the density of anhydrous LTG mixtures is slightly larger for w$_{\textrm{G}}$ = 5~\% than for w$_{\textrm{G}}$ = 0~\%
at temperatures below $\sim$ 400~K (Figure~\ref{density_glass_transition}a). To understand the origin of this behavior, we have investigated
the influence of water and glycerol on the density and the free volume of LTG and TG systems at 300 K, that is, in their glassy state.

\newpage
\subsection{Molecular packing}

\begin{figure}[htbp]
\includegraphics[width=8cm,clip=true]{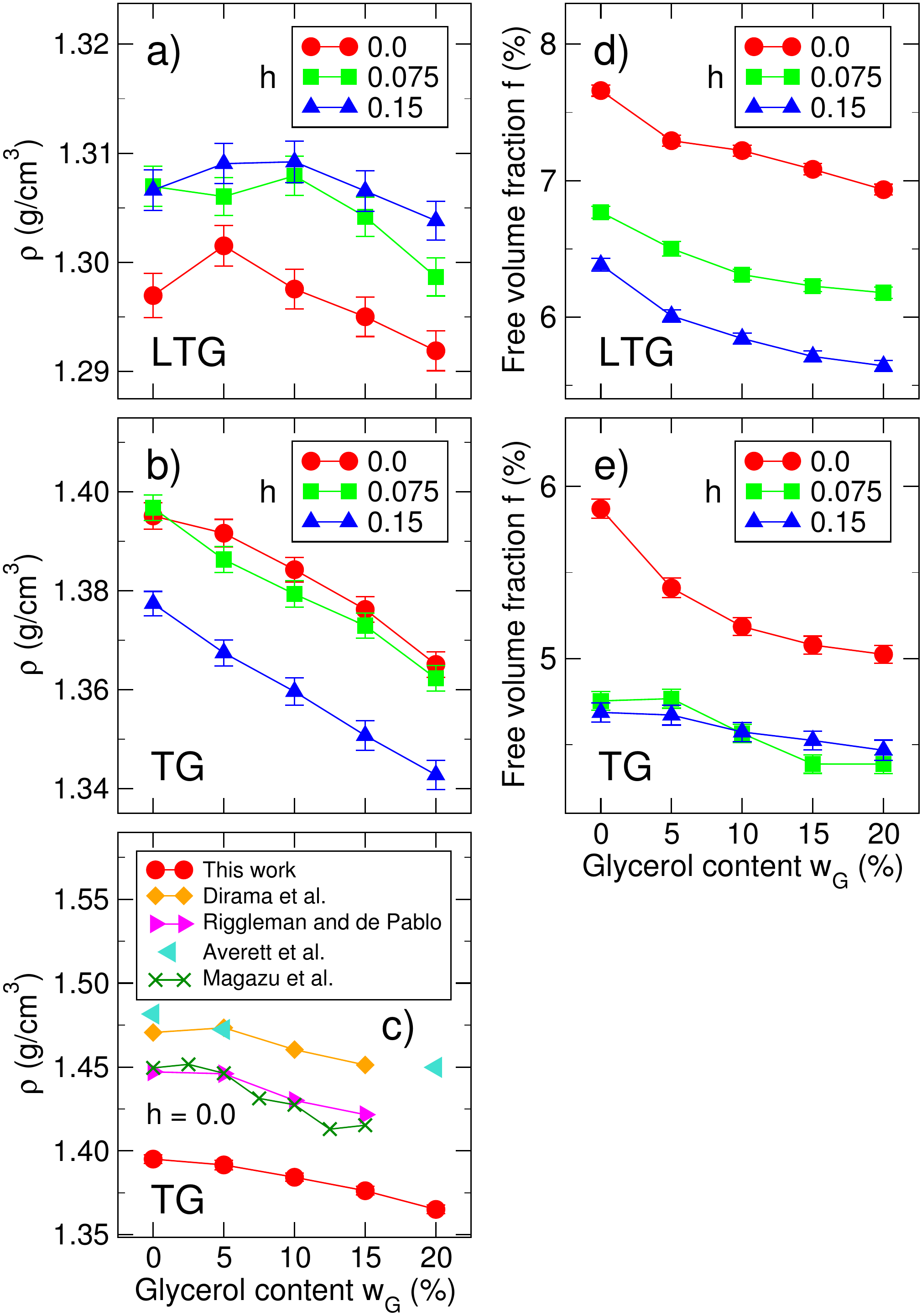}
\caption{\label{density_free_volume}
Left panel: Density, $\rho$, of the LTG (a) and TG (b) glasses as a function of the glycerol content, w$_{\textrm{G}}$,
and the hydration level, $h$, at 300~K. The densities of the anhydrous TG glasses obtained in this work are compared with those
determined in previous MD simulation studies~\cite{Dirama2005,Riggleman2008,Averett2012,Magazu2010} in (c).
Right panel: Free volume fraction, $f$, of the LTG (d) and TG (e) glasses as a function of the glycerol content, 
w$_{\textrm{G}}$, and the hydration level, $h$, at 300~K. $f$ denotes the ratio of the total free volume accessible to a probe radius of 0.5 \AA \/, 
$V_{f}$, to the total volume of the simulation box, V ($f$ = 100*$V_{f}$/$V$)~\cite{Limbach2008}. 
Rather similar results for $\rho$ and $f$ were obtained at 100 K (see Figure S3 in the Supporting Information).
}
\end{figure}

Figure~\ref{density_free_volume}a-b shows the densities, $\rho$, of the LTG and TG matrices at 300 K as a function of the
glycerol concentration, w$_{\textrm{G}}$, at various hydration levels, $h$. The dependence of $\rho$ on w$_{\textrm{G}}$ 
and $h$ is strikingly different whether lysozyme is present or not: whereas $\rho$ decreases monotonically 
when w$_{\textrm{G}}$ or $h$ increases in the TG mixtures (Figure~\ref{density_free_volume}b), 
small maxima systematically emerge for the LTG glasses at glycerol 
contents of 5-10 wt.~\% (Figure~\ref{density_free_volume}a), 
whatever the water content, $h$, considered (the largest density increase occurs in the anhydrous LTG mixture upon the addition 
of 5 \% of glycerol and is about 0.005 g.cm$^{\textrm{-3}}$). 
The decrease of density of TG mixtures with w$_{\textrm{G}}$ is expected when considering the smaller density of 
liquid glycerol ($\sim$ 1.26 g.cm$^{\textrm{-3}}$)~\cite{Bosart1928}
compared to that of amorphous trehalose ($\sim$ 1.50 g.cm$^{\textrm{-3}}$)~\cite{Imamura2008}.
It is consistent with that found by Averett~et al.~\cite{Averett2012} with the original OPLS force field for
trehalose and an optimized version for glycerol, even though the densities of our mixtures 
are downshifted by about 0.1 g.cm$^{\textrm{-3}}$ (Figure~\ref{density_free_volume}c). This difference probably stems from the CHARMM36 carbohydrate 
force field~\cite{Guvench2008,Guvench2009,Hatcher2009}, 
which underestimates the densities of concentrated solutions and crystals of carbohydrates~\cite{Guvench2009}
as well as the density of bulk glycerol~\cite{Averett2012}. In contrast, slight density maxima appear at w$_{\textrm{G}}$ of 5 and 2.5~\% in the MD 
studies of Dirama et al.~\cite{Dirama2005} and of Magazu~et~al.~\cite{Magazu2010}, 
respectively (Figure~\ref{density_free_volume}c), as well as at w$_{\textrm{G}}$ = 5~\% in that of Riggleman and de Pablo, 
for temperatures below $\sim$ 250-280~K (see Figure~1 in ref.~[\citen{Riggleman2008}]). The presence of a maximum in the dependence 
of the density of TG mixtures on w$_{\textrm{G}}$ thus relies on the force fields used in MD simulations 
to represent trehalose and glycerol. Experimental density data would help to capture such a subtle effect and to improve force fields accordingly.
Nonetheless, the distinct density behaviors of LTG and TG matrices in the present study reveal substantial differences in 
the relative packing efficiencies of lysozyme and trehalose molecules: excluded volume effects are 
much stronger in the presence of lysozyme than in its absence, owing to the globular shape and 
the much larger size of lysozyme in comparison with trehalose (radii of gyration of about 14.3 and 3.5 \AA \/, respectively), 
which obviously reduce its packing efficiency. As a consequence, 
the density of amorphous trehalose~\cite{Imamura2008} exceeds that of freeze-dried powders of proteins of various 
molecular weights - human growth hormone (22~kDa), bovine serum albumin (66~kDa), or immuno globulin G (150~kDa) - 
determined using gas pycnometry~\cite{Kikuchi2011}. This rationalizes why the densities of TG mixtures are systematically larger 
than those of LTG ones, at given glycerol and water contents. Besides, the increases in the density of LTG mixtures with 
w$_{\textrm{G}}$ and $h$ may be ascribed to the small sizes of water and glycerol, which can thereby access volumes 
inaccessible to the bigger trehalose (note that the mean molecular volumes occupied by water, glycerol, and trehalose are 
about 30, 120, and 380 \AA \/$^{\textrm{3}}$ in the bulk, assuming that their bulk densities are 1.0~\cite{Kell1967}, 1.26~\cite{Bosart1928}, and 
1.50 g.cm$^{\textrm{-3}}$~\cite{Imamura2008}, respectively). Interestingly, the increase of $\rho$ when $h$ increases from 0.0 to 0.075 is 
significantly larger than that found when w$_{\textrm{G}}$ raises from 0 to 5-10 \%, suggesting that, owing to its smaller size, 
water may enter regions at the protein-matrix interface that are sterically inaccessible to glycerol.\\

The non-monotonic dependence of $\rho$ on w$_{\textrm{G}}$ and $h$ in the LTG mixtures clearly suggests that
glycerol and water can act as antiplasticizers at low concentrations. MD simulations of a coarse-grained polymer indeed 
indicated that antiplasticization is linked to enhanced packing in the glassy state~\cite{Riggleman2010}.
To further study the influence of glycerol and water on the molecular packing of LTG and TG matrices, 
we then determined their free volume fraction, $f$~\cite{Limbach2008}, in a similar way as in ref.~[\citen{Molinero2003}] 
(Figure~\ref{density_free_volume}d-e). $f$ steadily decreases upon addition of glycerol, in line with the 
positron annihilation lifetime spectroscopy (PALS) and Fourier transform infrared (FTIR) study of Roussenova~et al. on maltodextrin-glycerol
amorphous matrices, which revealed that the addition of glycerol at concentrations up to 20 wt. \% reduces non-linearly 
the average molecular hole size, $v_{h}$, thereby enhancing molecular
packing~\cite{Roussenova2014}. $f$ also diminishes with $h$ in the LTG and TG matrices, which seems consistent
with the decrease of hole volume found for glassy maltopolymer-maltose and maltodextrin-glycerol matrices upon sorption
of low contents of water (up to weight fractions of $\sim$ 0.04-0.08)~\cite{Townrow2010,Roussenova2010,Roussenova2014,Ubbink2016}.
The larger $f$ values for LTG matrices in comparison with TG ones may be ascribed to a greater structural disorder
when mixing molecules of significantly different sizes and also, to a smaller extent, to internal protein cavities.
This difference in molecular packing between LTG and TG mixtures probably explains why $f$ keeps decreasing when $h$ 
increases from 0.075 to 0.15 for a given glycerol content in the LTG matrices, while $f$ hardly changes with $h$ in the TG matrices 
for the same increase of water content.

\newpage

\newpage

\begin{figure}[htbp]
\includegraphics[width=6cm,clip=true]{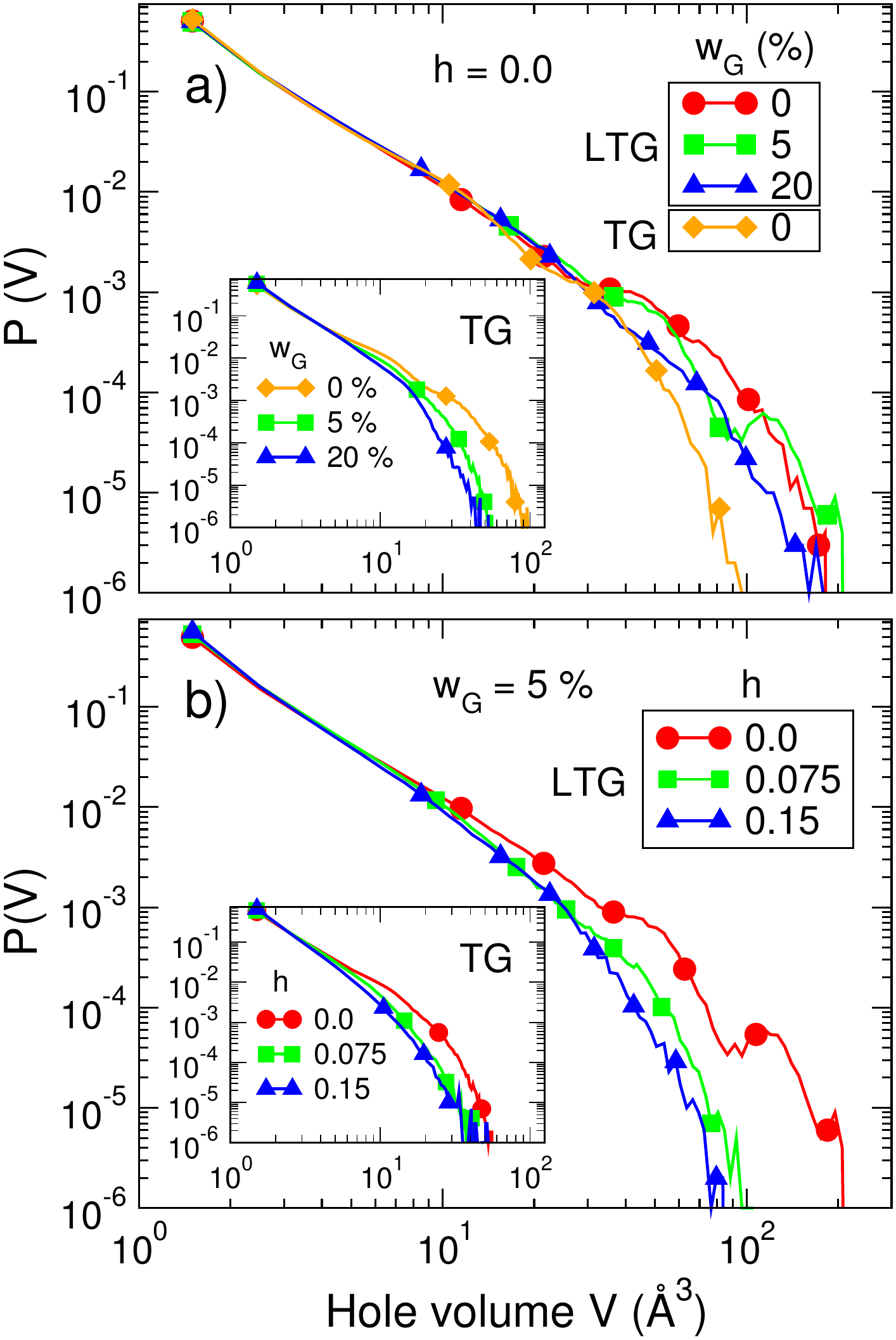}
\caption{\label{free_volume_distribution}
Probability distribution functions of the size of free volumes accessible to a probe radius of 1.0 \AA \/, P(V), at 300~K: 
(a) in the anhydrous LTG matrices at w$_{\textrm{G}}$ = 0, 5, and 20 \% and, for comparison, in the anhydrous TG matrix
at w$_{\textrm{G}}$ = 0 \%, (b) in the LTG matrices at w$_{\textrm{G}}$ = 5 \% and hydration levels, $h$, of 0.0,
0.075, and 0.15. The corresponding distributions for the TG matrices are shown in insets. Pictures on the spatial distributions
of free volume in representative configurations of LTG and TG matrices at 300~K are provided in Figure S4 in the Supporting Information.
}
\end{figure}

In order to get a more comprehensive insight into the molecular packing of LTG and TG
mixtures, we also computed the distribution of hole volumes in these systems
(Figure~\ref{free_volume_distribution}).
The addition of glycerol globally tends to reduce the number of holes larger than $\sim$
30 \AA \/$^{\textrm{3}}$.
This effect clearly appears in the TG matrices, in which the absence of lysozyme may
facilitate the packing between trehalose and glycerol molecules. In contrast, the number of holes bigger than
$\sim$ 200 \AA \/$^{\textrm{3}}$ is slightly larger at w$_{\textrm{G}}$ = 5 \% than at w$_{\textrm{G}}$ = 0 \% for the
anhydrous LTG matrix, which may indicate a lack of equilibration of LTG matrices. Moreover, in agreement with the results
shown in Figure~\ref{density_free_volume}d-e, the shift of distributions towards smaller hole volumes is more pronounced
upon addition of water than when adding glycerol. This result can obviously be ascribed to the smaller size of water,
which can intercalate more easily between lysozyme, trehalose, and glycerol molecules. Note, however, that the shift to low
hole volumes cannot be simply explained by a "hole filling" mechanism, as discussed by Roussenova~et al.
in ref.~[\citen{Roussenova2010}] (see also a short discussion and Figure S5 in the Supporting Information).

\newpage

%%\newpage

\begin{figure}[htbp]
\includegraphics[width=8cm,clip=true]{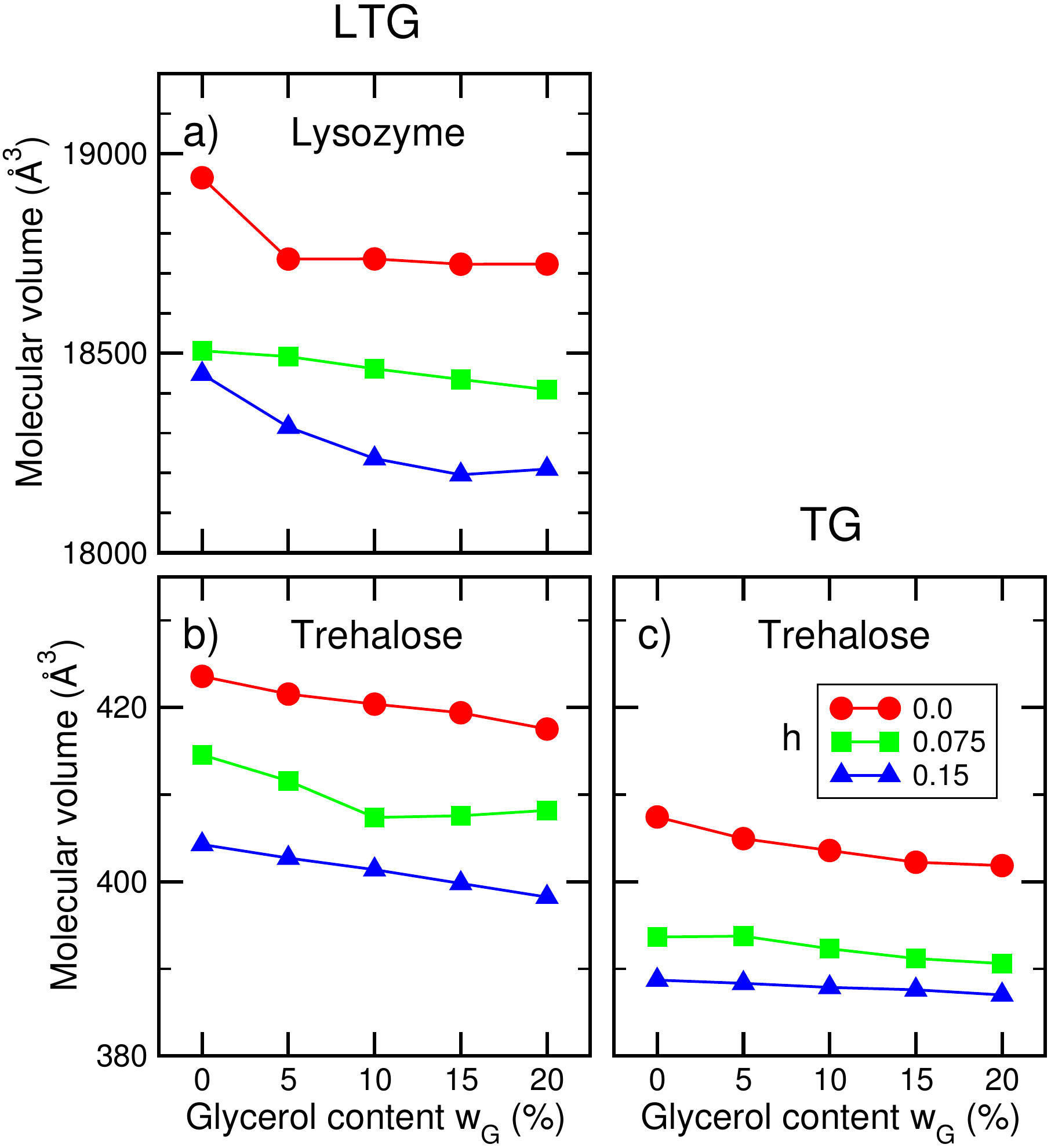}
\caption{\label{molecular_volume}
Average molecular volumes, $\overline{\textrm{V}}$, of lysozyme (a) and trehalose (b) in the LTG glasses at 300~K, as a function of the glycerol 
content, w$_{\textrm{G}}$, and for the different hydration levels, $h$. $\overline{\textrm{V}}$ of trehalose in the TG glasses at 300~K are 
displayed in (c). $\overline{\textrm{V}}$ were computed
using Voronoi tessellation as performed by the program of Gerstein~{et al.}~\cite{Gerstein1995}. For clarity,
the standard deviations around mean values are not shown as error bars, but are given in Table~S3 in the Supporting Information.
}
\end{figure}

As a way to describe how glycerol and water modify the local free volume of lysozyme and trehalose in the LTG and TG glasses,
we then determined their average molecular volumes, $\overline{\textrm{V}}$, using Voronoi tessellation~\cite{Gerstein1995} 
(Figure~\ref{molecular_volume}).
$\overline{\textrm{V}}$ of lysozyme and trehalose tend to decrease when w$_{\textrm{G}}$ and/or $h$ increase in the simulated
systems. The decreases are particularly steep in the anhydrous LTG matrix when w$_{\textrm{G}}$ increases from 0 to 5 wt. \% 
($\sim$ -1 \%) or when $h$ increases from 0.0 to 0.075 ($\sim$ -2 \%). These results definitely indicate that glycerol and water
enhance the packing of lysozyme and trehalose molecules, consistent with Figures~\ref{density_free_volume} and \ref{free_volume_distribution}.
Nevertheless, the molecular volumes of lysozyme and trehalose in LTG and TG matrices exceed those found experimentally 
in dilute aqueous solutions at room temperature~\cite{Makhatadze1990,Miller1997} by $\sim$ 7-11 \% for lysozyme 
in LTG systems and by $\sim$ 14-21 \% and $\sim$ 11-16 \% for trehalose in LTG and TG matrices, respectively. 
The partial molar volumes of lysozyme and trehalose were indeed determined to be 
10280 cm$^{\textrm{3}}$.mol$^{\textrm{-1}}$ $\approx$ 17070 \AA \/$^{\textrm{3}}$ for lysozyme~\cite{Makhatadze1990} and 
$\sim$ 210 cm$^{\textrm{3}}$.mol$^{\textrm{-1}}$ $\approx$ 350 \AA \/$^{\textrm{3}}$ for trehalose~\cite{Miller1997}.
These differences probably stem in part from the underestimated densities of simulated systems. But, they could also indicate that both 
lysozyme and trehalose are not as well solvated in the studied glasses as they are in dilute aqueous solutions,
owing to significant excluded volume effects that do not allow lysozyme to interact as intimately with trehalose and glycerol 
as it interacts with water. Similarly, the larger $\overline{\textrm{V}}$ of trehalose in the LTG matrices than in the TG ones 
at given water and glycerol contents shows that trehalose cannot interact as efficiently with lysozyme as it interacts with itself.

\newpage

\subsection{Mean-square displacements}

\begin{figure}[htbp]
\includegraphics[width=6.5cm,clip=true]{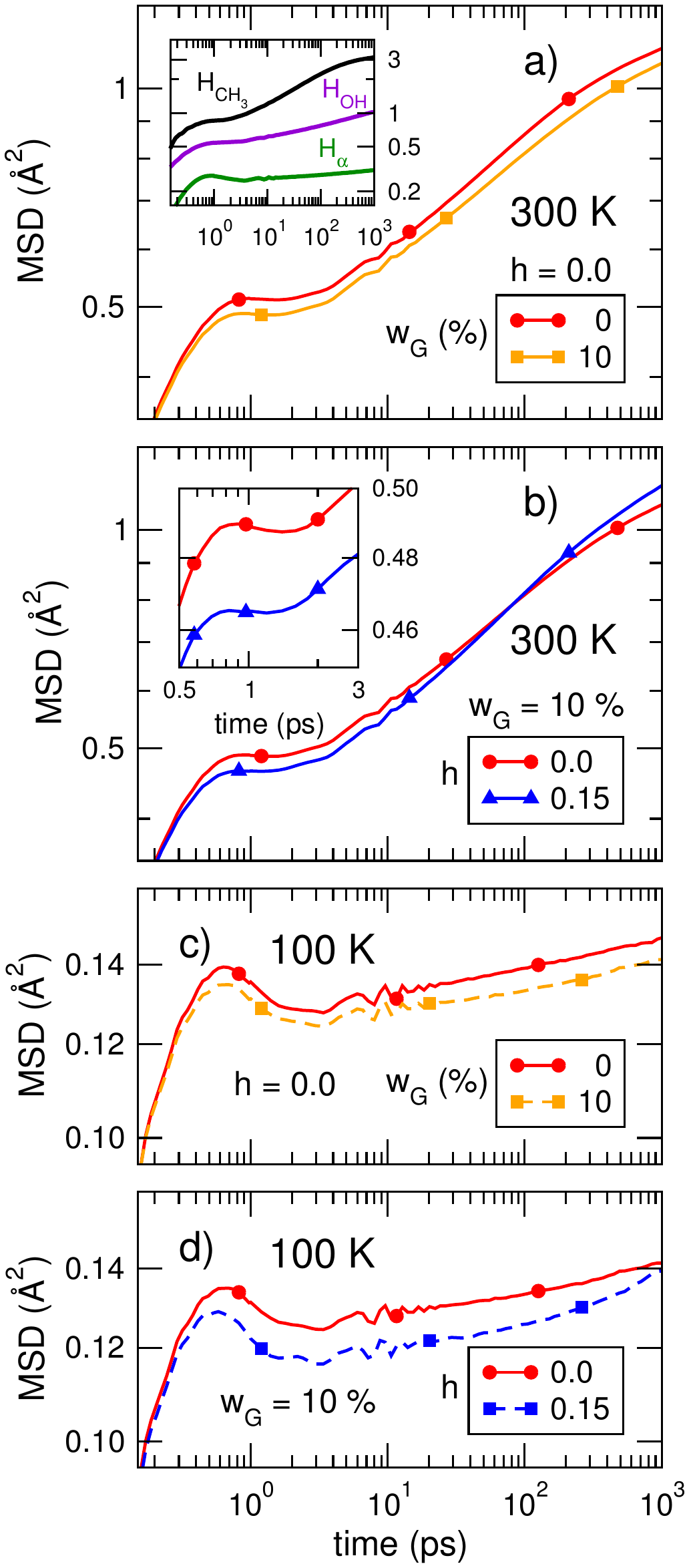}
\caption{\label{msd_time}
Time dependence of the mean-square displacement (MSD) of the hydrogens of lysozyme in LTG matrices~:
(a) at $h$ = 0.0 and w$_{\textrm{G}}$ = 0 or 10~\%, (b) at w$_{\textrm{G}}$ = 10~\% and
$h$ = 0.0 or 0.15. The corresponding MSDs at 100~K are shown in (c) and (d), respectively.
The inset of (a) shows the MSDs of three kinds of hydrogens of lysozyme in the anhydrous LTG matrix
at w$_{\textrm{G}}$ = 0~\%~: (i) methyl groups, H$_{\textrm{CH}_3}$ (from ALA, ILE, LEU, MET, THR, and VAL residues), 
(ii) hydroxyl groups, H$_{\textrm{OH}}$ (from SER, THR, and TYR residues),
and (iii) H$_{\alpha}$ backbone atoms of lysozyme. The inset of (b) shows a zoom on the 0.5-3~ps time range 
to make clearer the decrease of the MSD of lysozyme induced by the addition of water.
The time dependences of the MSDs of lysozyme, trehalose, glycerol, and water in LTG and TG glasses 
at w$_{\textrm{G}}$ = 10~\% and $h$ = 0.15 are provided in Figure~S6 in the Supporting Information.
}
\end{figure}

The improved molecular packing of LTG and TG glasses induced by the addition of glycerol and/or water
(Figures~\ref{density_free_volume}-\ref{molecular_volume}) suggests that glycerol and water also
decrease the short-time scale MSD of LTG and TG mixtures, since a relationship between the free volume and the Debye-Waller (DW)
factor was found for glass-forming systems (see, for instance, ref.~[\citen{Starr2002,Ottochian2011}]).
Figure~\ref{msd_time}a shows the time dependence of the MSD of the hydrogens of lysozyme in the anhydrous LTG matrices with 
w$_{\textrm{G}}$ = 0 and 10~\%. The plateau that appears at a time of 1-2~ps reflects vibrational and rattling motions of atoms within the 
cage formed by their neighbors~\cite{Hong2011}, while the steep increase that follows arises from anharmonic motions. 
Hong~et al. identified three classes of motions for the nonexchangeable hydrogen atoms of lysozyme~: "localized diffusion", 
"methyl group rotations", and "jumps"~\cite{Hong2011}. 
The distinct time dependences of the MSDs of lysozyme's methyl, hydroxyl, and H$_{\alpha}$ backbone hydrogen atoms illustrate
such a great heterogeneity of motions (inset of Figure~\ref{msd_time}a). The detailed analysis of such motions is beyond the scope 
of this study, so that we will only consider and discuss the influence of water and glycerol on the average MSDs of lysozyme's 
and trehalose's hydrogen atoms in the following.
Figure~\ref{msd_time}a evidences that the presence of glycerol can decrease the MSD of lysozyme in the ps-ns time scale.
In contrast, water is found to slightly decrease the MSD of lysozyme in the plateau regime ($\sim$~1~ps), whereas it 
increases it at longer time scales ($> \sim$~100~ps for the LTG matrix with w$_{\textrm{G}}$ = 10~\%
and $h$~=~0.15, Figure~\ref{msd_time}b). Figure~\ref{msd_time}c-d shows the corresponding MSDs at 100~K, at which 
the thermal activation of methyl group rotations~\cite{Roh2005,Hong2013} and of jumps is inhibited~\cite{Hong2013}.
At such a low temperature, both glycerol (Figure~\ref{msd_time}c) and water (Figure~\ref{msd_time}d) reduce the MSD of lysozyme 
from the ps to the ns time scale.

\begin{figure}[htbp]
\includegraphics[width=14cm,clip=true]{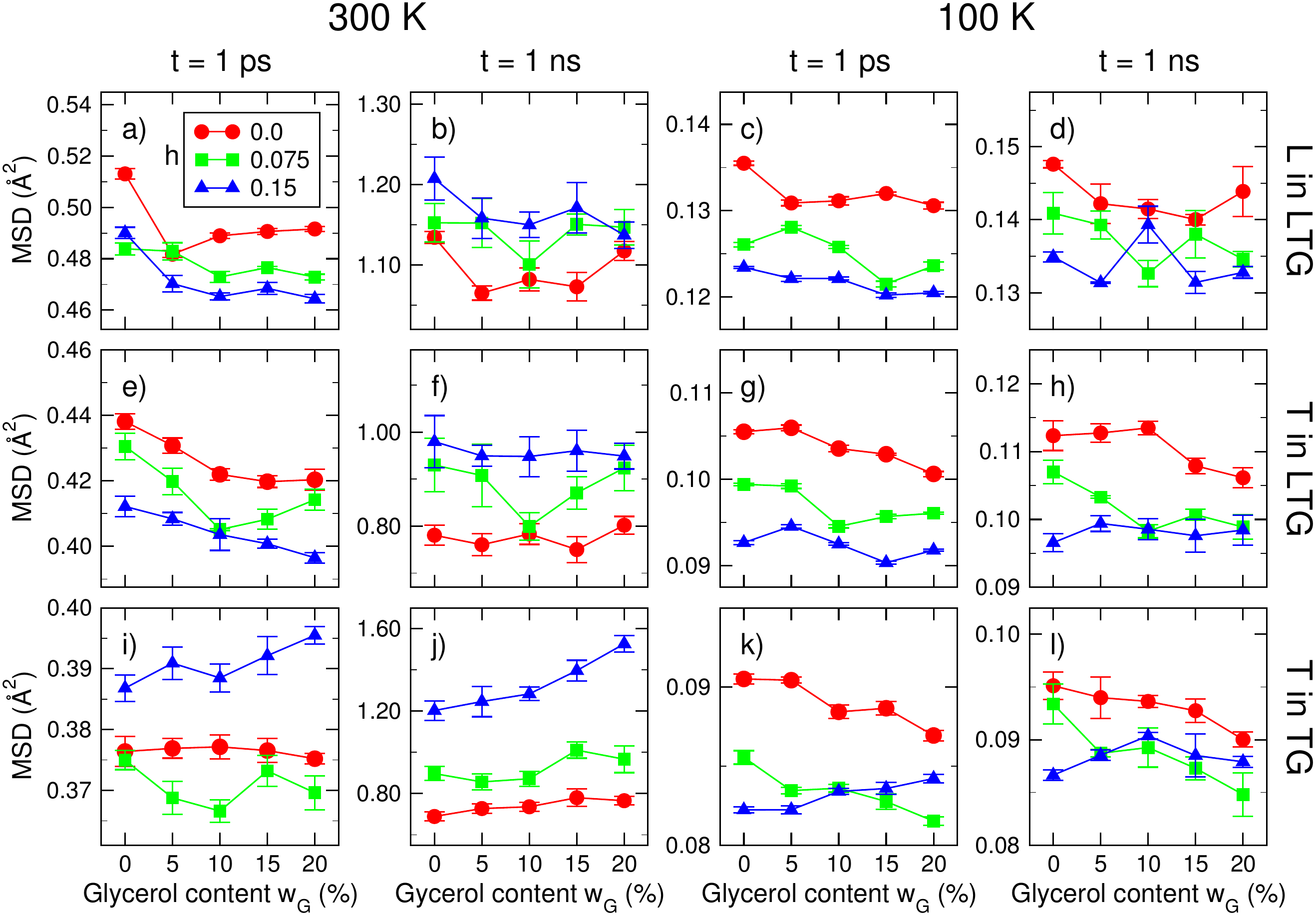}
\caption{
\label{msd}
Mean-square displacements (MSDs) of the hydrogens of lysozyme and trehalose in LTG and TG glasses 
as a function of the glycerol content, w$_{\textrm{G}}$, and for the different hydration levels, $h$~:
(a-d) L in LTG, (e-h) T in LTG, and (i-l) T in TG matrices. MSDs were determined at two time scales, 1~ps and 1~ns, 
for the two temperatures considered, 300~K (left two panels) and 100~K (right two panels). Error bars correspond 
to standard deviations from mean values determined by splitting trajectories into 5 sub-trajectories for all MSDs 
at 300~K and for MSDs at 100~K at the ps time scale, while 3 sub-trajectories were considered for MSDs at 100~K at the 
ns time scale.
}
\end{figure}

In order to investigate in a systematic and straightforward manner the effects of water and glycerol on the MSDs 
of lysozyme and trehalose in LTG and TG matrices, we then compared MSDs at two time scales, 1~ps and 1~ns, both 
at 300~K and at 100~K (Figure~\ref{msd}). In LTG matrices, 
the addition of glycerol tends to lower the MSDs of L and T at both 
time scales and temperatures (with a few exceptions, where the MSDs increase moderately, usually within few standard deviations), thereby 
indicating that glycerol may act as an \textit{antiplasticizer} on lysozyme and trehalose fast motions in this concentration range. 
This is in fair agreement with the neutron scattering study of Tsai et al., who found that the addition of glycerol 
lowers the amplitude of motions of Lysozyme/Glycerol mixtures (80~wt.~\%/20~wt.~\%) with respect to those of dry lysozyme at temperatures 
below the so-called \textit{dynamic transition temperature}, $T_{d}$, of 330~K for this mixture~\cite{Tsai2000}.
Besides, antiplasticization of trehalose motions by glycerol is absent in TG matrices at $h$ = 0.15 (Figure~\ref{msd}i-l). It emerges 
only at lower water contents, rather modestly at 300~K (Figure~\ref{msd}i-j), but much more clearly at 100 K (Figure~\ref{msd}k-l). 
This result appears at variance with the neutron scattering data of Cicerone and Soles~\cite{Cicerone2004}, which evidenced a 
significant decrease of the MSD of trehalose in TG mixtures at w$_{\textrm{G}}$ = 5 \% over a broad temperature range 
($\sim$ 100-350 K). But, the modest decrease of MSD induced by glycerol in TG matrices fairly agrees with later studies 
by Riggleman and de Pablo~\cite{Riggleman2008}, by Magazu~et~al.~\cite{Magazu2010}, and by Averett~et~al.~\cite{Averett2012}. 
Furthermore, the strong temperature dependence of 
the antiplasticizing effect of glycerol on trehalose motions appears consistent
with that experimentally found by Anopchenko et al. in TG matrices from dielectric relaxation spectroscopy~\cite{Anopchenko2006,Obrzut2010}. 
Besides, the influence of water on the fast motions of 
lysozyme and trehalose in LTG and TG matrices is also complex and strongly depends on the time scale 
and temperature considered: at 300~K, water diminishes the MSDs of L and T at the ps time scale 
(only at $h$=0.075 in TG matrices, Figure~\ref{msd}a,e,i), but raises them at the ns time scale (Figure~\ref{msd}b,f,j). 
In contrast, water globally tends to decrease 
the MSDs of both L and T at the ns time scale at 100 K (Figure~\ref{msd}d,h,l), thereby demonstrating that it stiffens vibrational, 
(quasi-)harmonic motions at low temperatures. 
These results corroborate those of Nickels et al. from neutron scattering measurements on green fluorescent protein 
(GFP)~\cite{Nickels2012}, which showed that the MSD of GFP at a hydration level, $h$, of 0.4 g(D$_2$O)/g(protein)
is either enhanced or reduced with respect to that of the dry protein for temperatures larger or lower than
$\sim$ 240 K, respectively~\cite{Nickels2012}. They are also consistent with those of Hong~et al., who showed that the effective
force constant derived from MD simulation as well as the frequency of longitudinal sound waves, $\nu_{L}$, determined from 
Brillouin light scattering experiments are larger for hydrated than for dry GFP at temperatures below $\sim$ 180~K~\cite{Hong2013}.
Analogous temperature-dependent effects of water on the MSD of RNase A were also found by Tarek and Tobias from MD 
simulation~\cite{Tarek2008}.

Finally, we checked whether the changes of MSD of LTG and TG matrices upon the addition of glycerol or water 
are related to changes in density. For this purpose, we assumed that the MSD at a given glycerol or water content, $x$
($x$ = w$_{\textrm{G}}$ or $h$), may be written as $MSD(x) = MSD(0)+\alpha.(\rho(x)-\rho(0))$, where $MSD(0)$ and $\rho(0)$
correspond to the MSD and density in the absence of the considered diluent, and $\alpha$ is a constant 
(see further details in the Supporting Information). Although very simple, this phenomenological equation is able to 
qualitatively account for the minima in the MSDs of LTG matrices or for the increase of the MSD of TG ones observed at the ps 
time scale upon addition of glycerol at 300~K (see Figure~S7a,e in the Supporting Information). Moreover, it clearly relates the 
antiplasticizing effect of water on the ps-time scale motions of LTG matrices at 100 K to the corresponding increase of density 
(see Figure~S8c in the Supporting Information). Nevertheless, a much deeper analysis would be required to explain on
theoretical grounds how MSD depends on density in such systems.

In the following, we will describe thoroughly the intermolecular hydrogen bonds formed between species 
to clarify the complex dependences of the MSDs of lysozyme and trehalose on the concentrations of glycerol and water,
on temperature, and on the time scale considered.

\newpage

\subsection{Hydrogen bonds}

\begin{figure}[htbp]
\includegraphics[width=14cm,clip=true]{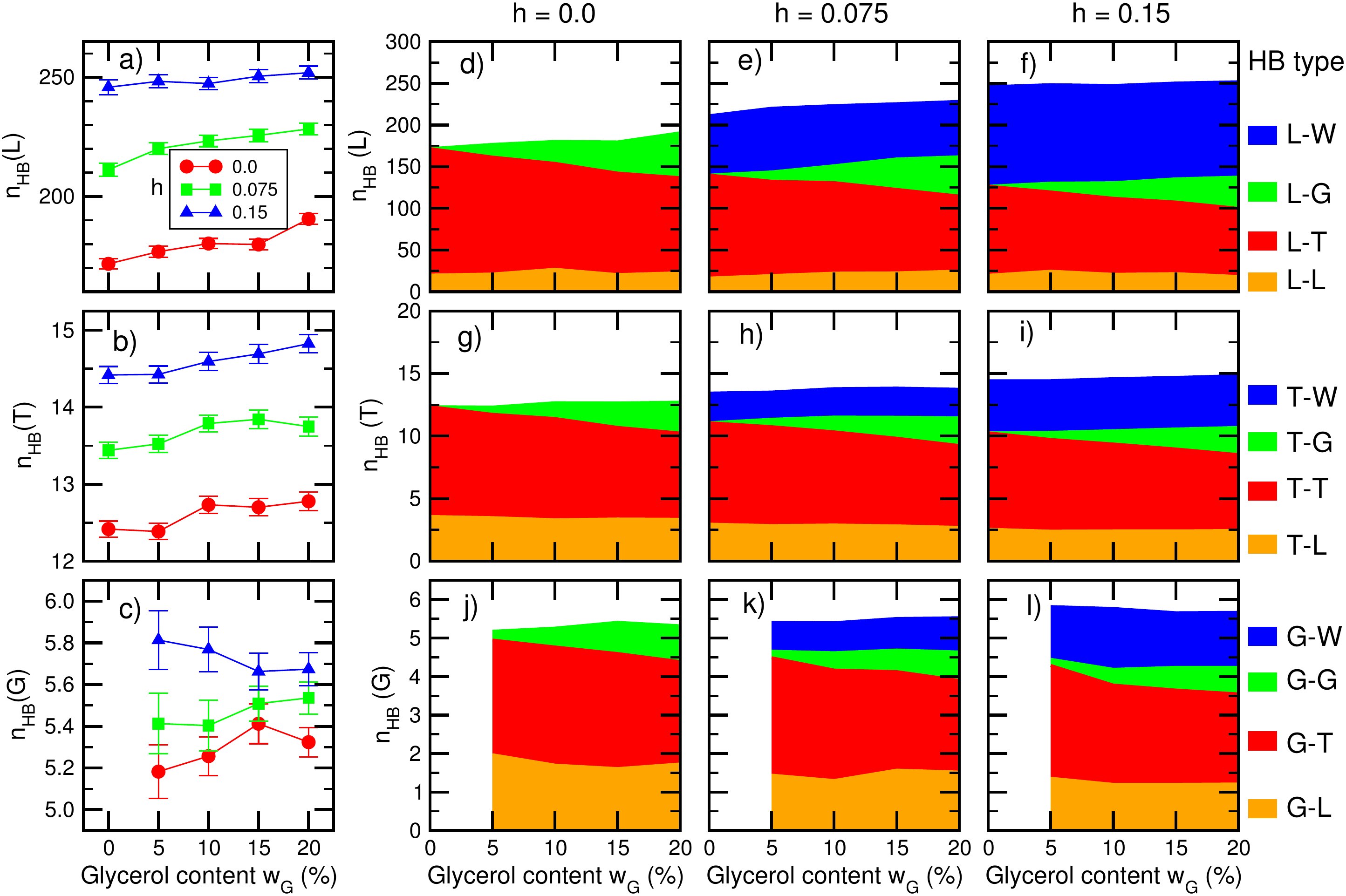}
\caption{\label{nb_HBs}
Total numbers of intermolecular HBs, $n_{HB}$, formed by lysozyme, $n_{HB}$(L) (a), trehalose, $n_{HB}$(T) (b), and 
glycerol, $n_{HB}$(G) (c) in the LTG matrices as a function of the glycerol content, w$_{\textrm{G}}$, for the different 
hydration levels, $h$, at 300~K. The decompositions of $n_{HB}$(L), $n_{HB}$(T), and $n_{HB}$(G)
into contributions from HBs involving each species are shown in d-f, g-i, and j-l, respectively (the corresponding
numerical values are provided in Table~S4 in the Supporting Information).
Moreover, the $n_{HB}$ of lysozyme and trehalose in LTG matrices at 100 K, as well as the $n_{HB}$ of trehalose in TG matrices
at 300 K and at 100 K are shown in Figure S9 in the Supporting Information.
}
\end{figure}

Previous studies evidenced that the stability of proteins in the solid state is improved
when they form numerous HBs with excipients (see \textit{e.g.} ref.~[\citen{Allison1999}] and reviews in 
ref. [\citen{Wang2000,Manning2010,Ohtake2011}]). It has been proposed that these HBs substitute for
those that proteins form in aqueous solution with their hydration water and, thereby, maintain the native 
structure and function of proteins in the dry state (\textit{water replacement} hypothesis~\cite{Crowe1984b}).
Furthermore, the antiplasticizing effect of glycerol on trehalose has been correlated with the formation 
of T-G HBs in previous MD studies~\cite{Dirama2005,Curtis2006}, even though antiplasticization has also 
been observed in model systems that do not form HBs~\cite{Simmons2011}. Thus, we characterized thoroughly 
the distribution of HBs between the different species, as well as their geometry and dynamics, to deeply understand the complex 
influences of glycerol and water on the fast motions of lysozyme and trehalose observed in Figures~\ref{msd_time}-\ref{msd}. 
For this purpose, we 
determined the intermolecular HBs formed between the different species following the same geometric 
criterion that we used previously~\cite{Lerbret2007,Lerbret2012}~: a hydrogen bond between a pair of 
donor, D, and acceptor, A, atoms was considered to exist if the D$\cdot\cdot\cdot$A distance was less than 
3.4 \AA \ and if the D-H$\cdot\cdot\cdot$A angle was larger than $120\,^{\circ}$. 
Figure~\ref{nb_HBs}a-c shows the total numbers of intermolecular HBs, $n_{\textrm{HB}}$, that lysozyme, trehalose, and 
glycerol form in the LTG mixtures, as a function of the glycerol content, w$_{\textrm{G}}$, and the hydration level, $h$. 
The $n_{\textrm{HB}}$ of both lysozyme and trehalose increase much more with $h$ than with w$_{\textrm{G}}$.
This difference obviously stems from the smaller size and simpler topology of water compared to glycerol, which 
allow a more intimate interaction with lysozyme and trehalose. It is fully consistent with the larger decreases
of the molecular volumes $\overline{V}$ of L and T found when adding water than upon addition of glycerol (Figure~\ref{molecular_volume}). 
Moreover, the $n_{\textrm{HB}}$ of glycerol increases systematically with $h$ and tends to increase with 
w$_{\textrm{G}}$ at $h$ = 0.0 and 0.075. Besides, Figure~\ref{nb_HBs}a demonstrates that lysozyme 
remains only partially hydrated in the LTG matrices, as we previously found for LT matrices in ref.~[\citen{Lerbret2012}].
This indicates that trehalose and glycerol are unable to fully replace the HBs that lysozyme forms in dilute aqueous solutions 
(we found that lysozyme forms about 330 HBs with water in ref.~[\citen{Lerbret2007}]). Therefore, 
the \textit{water replacement hypothesis}~\cite{Crowe1984b} is only valid to some extent. Furthermore, at 
given w$_{\textrm{G}}$ and $h$, trehalose forms about two HBs less in LTG matrices than in TG ones (see Figure S9 in the Supporting Information). 
This can be ascribed to the disorganization of the molecular packing of trehalose molecules induced by
lysozyme and it corroborates the larger molecular volumes $\overline{V}$ of trehalose in LTG matrices (Figure~\ref{molecular_volume}b-c). 
The roughness of the protein surface as well as excluded volume effects between neighboring proteins reduce 
efficient protein-sugar H-bonding interactions. This result appears compatible with the lower density and larger free
volume fraction of LTG systems compared to TG ones (Figure~\ref{density_free_volume}). The lower $n_{\textrm{HB}}$(T) in LTG
matrices is also consistent (i) with the results of Carpenter and Crowe, which showed that the capacity of carbohydrates 
for hydrogen bonding decreases when they are mixed with lysozyme~\cite{Carpenter1989}, and (ii) with those of 
Katayama~et al., who found a positive excess enthalpy when mixing RNase A with trehalose or sucrose, 
thereby suggesting weaker or fewer HBs in the protein/sugar mixtures than in the two pure components~\cite{Katayama2009}.

The decomposition of $n_{\textrm{HB}}$ into contributions from each species provides useful information 
(Figure~\ref{nb_HBs}d-l and Table~S4 in the Supporting Information). First, the numbers 
of L-T and L-W HBs clearly overwhelm that of L-L HBs. This may explain why protein aggregation during storage is
reduced in presence of trehalose, which physically separate proteins from each other~\cite{Wang2005,Perez2010}.
Moreover, Figure~\ref{nb_HBs}d-f reveals that the increase of $n_{\textrm{HB}}$(L) with w$_{\textrm{G}}$ and $h$ stems
from the substitution of L-T HBs by L-G and/or L-W ones. Similarly, the $n_{\textrm{HB}}$ of trehalose in LTG and TG
matrices increase much more with $h$ than with w$_{\textrm{G}}$ (see Figure S9 in the Supporting Information), 
which obviously stems from the more significant substitution of
T-T HBs by T-W HBs than by T-G ones (see Tables~S4 and S5 in the Supporting Information).
In addition, Figure~\ref{nb_HBs}d-f and Table~S4 prove 
that trehalose, glycerol, and water are not homogeneously distributed around lysozyme in LTG matrices. Indeed, 
L-W HBs represent about 45-49 \% (depending on w$_{\textrm{G}}$) of $n_{\textrm{HB}}$(L) at $h$ = 0.15, even though 
water amounts for only $\sim$ 15 \% of the total mass of the solvent in these systems. Therefore, lysozyme interacts 
preferentially with water rather than with trehalose. Furthermore, the proportion of HBs involving lysozyme 
(L-L, T-L, G-L, and W-L) increases when the size of species decreases, thereby evidencing excluded volume effects: 
L-L interactions are much less likely than W-L ones, since water may interact intimately with lysozyme.
Accordingly, trehalose and glycerol form much more HBs with other trehalose and/or glycerol molecules than with
lysozyme.

\newpage

\subsubsection{Preferential interactions}

\begin{figure}[htbp]
\includegraphics[height=7.5cm,clip=true]{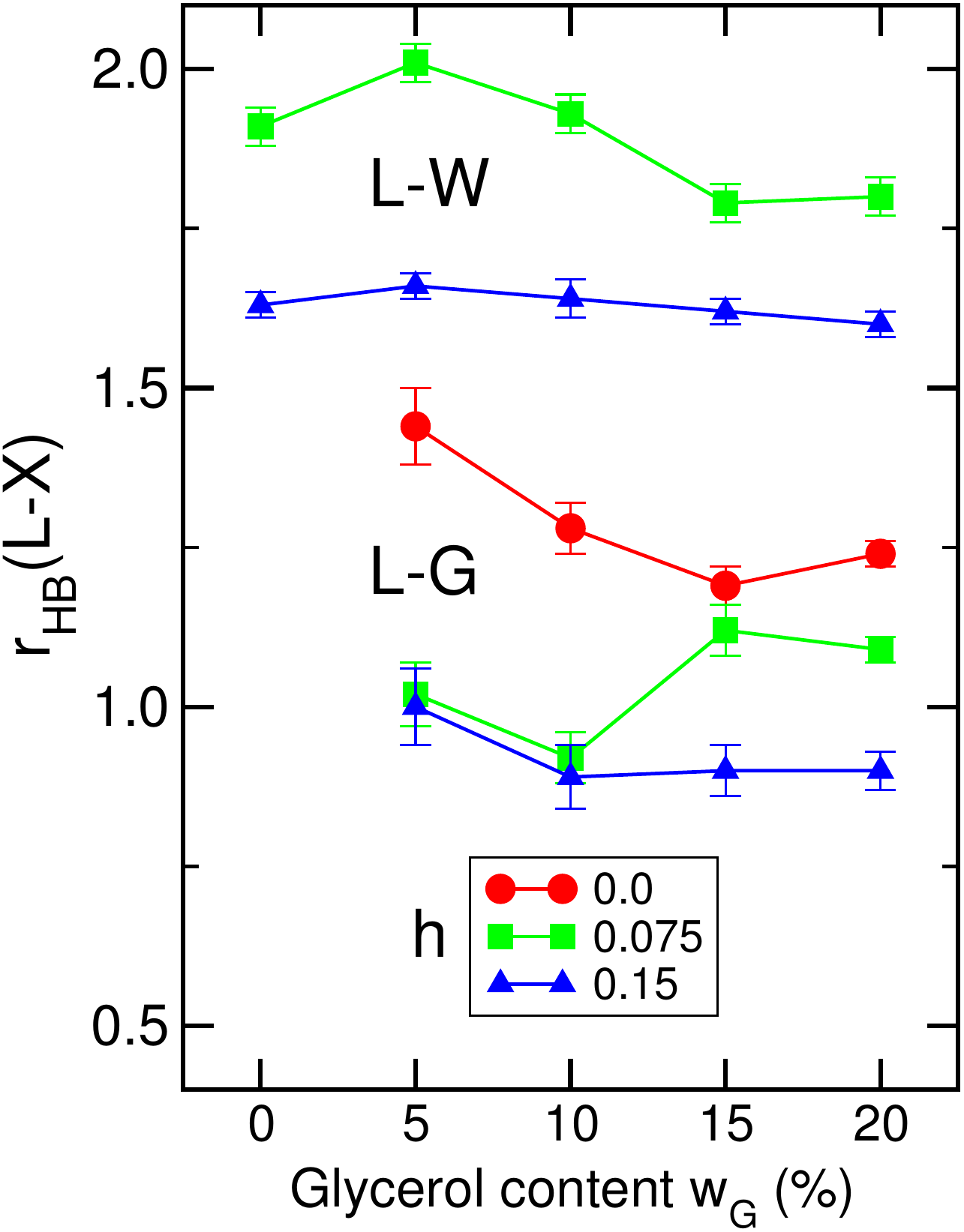}
\caption{\label{pref_int}
Ratio $r_{\textrm{HB}}$(L-X) for L-G and L-W HBs as a function of the glycerol content, w$_{\textrm{G}}$, for different 
hydration levels, $h$, at 300~K. $r_{\textrm{HB}}$(L-X)=$f_{\textrm{HB}}$(L-X)/$f_{\textrm{HB,theor.}}$(L-X), where 
$f_{\textrm{HB}}$(L-X)=$n_{\textrm{HB}}(L-X)$/[$n_{\textrm{HB}}$(L-T)+$n_{\textrm{HB}}$(L-G)+$n_{\textrm{HB}}$(L-W)],
and X stands for the considered species (X = T, G, or W). 
$f_{\textrm{HB,theor.}}$(L-X) denotes the fraction of HBs that species X (trehalose, glycerol, or water) 
could form with lysozyme assuming that each hydroxyl group of trehalose and glycerol may form three HBs with lysozyme 
(one as donor and two as acceptors) and each water molecule four (two as donors and two as acceptors). 
By definition, $r_{\textrm{HB}}$(L-X) indicates whether species X forms an excess 
($r_{\textrm{HB}}$(L-X)$>$1) or a lack ($r_{\textrm{HB}}$(L-X)$<$1) of HBs with lysozyme. 
}
\end{figure}

Given that Figure~\ref{nb_HBs}d-l suggests that trehalose, glycerol, and water have different affinities for lysozyme, 
we characterized these differences quantitatively by determining the fractions of HBs formed by trehalose, 
glycerol, and water with lysozyme, 
$f_{\textrm{HB}}$(L-X)=$n_{\textrm{HB}}(L-X)$/[$n_{\textrm{HB}}$(L-T)+$n_{\textrm{HB}}$(L-G)+$n_{\textrm{HB}}$(L-W)],
where X stands for the considered species (X = T, G, or W). We then normalized these fractions by the fractions
of HBs, $f_{\textrm{HB,theor.}}$(L-X)=$n_{\textrm{HB,theor.}}(L-X)$/[$n_{\textrm{HB,theor.}}$(L-T)+$n_{\textrm{HB,theor.}}$(L-G)+$n_{\textrm{HB,theor.}}$(L-W)], 
that species X (trehalose, glycerol, or water) could form with lysozyme assuming that
each hydroxyl group of trehalose and glycerol may form three HBs with lysozyme (one as donor and two as acceptors) and
each water molecule four (two as donors and two as acceptors). 
Under this hypothesis, $n_{\textrm{HB,theor.}}$(L-T)=3*8*N$_{T}$, $n_{\textrm{HB,theor.}}$(L-G)=3*3*N$_{G}$, 
and $n_{\textrm{HB,theor.}}$(L-W)=4*N$_{W}$, where N$_{T}$, N$_{G}$, and N$_{W}$ denote the numbers of trehalose, glycerol,
and water molecules, respectively, in the simulation box (see Table~S1 in the Supporting Information).
The normalized ratio 
$r_{\textrm{HB}}$(L-X)=$f_{\textrm{HB}}$(L-X)/$f_{\textrm{HB,theor.}}$(L-X) then indicates whether species X forms more HBs than
the one expected ($r_{\textrm{HB}}$(L-X)$>$1) or not ($r_{\textrm{HB}}$(L-X)$<$1). Figure~\ref{pref_int} displays the
ratios $r_{\textrm{HB}}$(L-G) and $r_{\textrm{HB}}$(L-W) for the different simulated LTG systems. In the absence of water, 
glycerol interacts preferentially with lysozyme in comparison with trehalose (an excess from about 40 \% for 
w$_{\textrm{G}}$ = 5 \% down to $\sim$ 20-25 \% for w$_{\textrm{G}}$ $\geq$ 15 \% is observed for L-G HBs). This definitely shows that glycerol is not 
homogeneously distributed in the LTG matrices, but rather that it is found slightly preferentially in the vicinity of the 
protein surface. This result can be ascribed to the smaller size of glycerol, which can therefore fit in empty spaces at 
the protein-matrix interface from which the bigger trehalose is size-excluded, and it explains the decrease of 
$\overline{V}$ of lysozyme when w$_{\textrm{G}}$ increases (Figure~\ref{molecular_volume}a). 
Upon addition of water ($h$ = 0.075-0.15), $r_{\textrm{HB}}$(L-G) decreases significantly,
so that glycerol forms within 10-15~\% of the number of HBs with lysozyme that one would expect. 
Conversely, water forms 
a great excess of HBs with lysozyme, consistent with the \textit{preferential hydration} hypothesis~\cite{Timasheff2002}
in dilute or semi-dilute solutions and with the \textit{water entrapment} hypothesis~\cite{Belton1994} in the solid 
state. This result corroborates those from previous MD studies on lysozyme and myoglobin in aqueous carbohydrate solutions 
and concentrated matrices~\cite{Cottone2002,Cottone2005,Lerbret2007,Lerbret2012,Corradini2013}.
In addition, $r_{\textrm{HB}}$(L-W) increases when $h$ decreases, in agreement with previous simulation results where 
the preferential hydration of lysozyme in aqueous disaccharide solutions was found to increase with the sugar 
concentration~\cite{Lerbret2007}. This supports the \textit{water anchorage hypothesis}~\cite{Cordone2005,Francia2008}, 
which suggests that the role of residual water is to \textit{anchor} the dynamics of proteins to that of the embedding matrix. 
It is worth mentioning that the preferential exclusion of trehalose from the surface of lysozyme with respect to glycerol probably 
originates solely from larger topological constraints and sterical hindrance effects on trehalose, rather than from 
differences in the affinity of their respective hydroxyl groups for lysozyme. Indeed, the hydroxyl groups of trehalose 
and glycerol share the same non-bonded parameters for electrostatic and van der Waals interactions in the CHARMM36 force 
field used in the current study~\cite{Guvench2008,Guvench2009,Hatcher2009}. 
In contrast, the preferential hydration of lysozyme probably does not exclusively results from 
topological and excluded volume effects, but could also arise from a greater chemical affinity of water for lysozyme, 
since, for instance, water exhibits a larger dipole moment than that of the hydroxyls of 
glycerol and trehalose molecules (the hydroxyl oxygen and hydrogen atoms of both trehalose and glycerol carry partial charges 
of -0.65~$e$ and +0.41~$e$, respectively, while the partial charges on water oxygen and hydrogen atoms are -0.8476~$e$
and +0.4238~$e$, respectively). Furthermore, it is important to keep in mind that the preferential hydration 
of lysozyme described above does not imply that trehalose 
remains far from the surface of lysozyme. About 80 to 90 \% of trehalose molecules actually form at least one HB with lysozyme 
in the LTG systems, depending on the water and glycerol contents considered (data not shown), and may thus be considered as interfacial. 
Therefore, distribution inhomogeneities of trehalose and glycerol around lysozyme are short-range, and preferential hydration 
in such concentrated protein matrices reflects their inability to form HBs with lysozyme as efficiently as water does rather than 
any phase separation.

\newpage 

Besides, the preferential interaction of lysozyme with water and, to a lower extent, with glycerol rather than with trehalose 
implies that glycerol and water can form \textit{bridges} between lysozyme and trehalose molecules (see Figures S10 and S11 in the 
Supporting Information). Interestingly, close to 80 \% of glycerol molecules are shared between lysozyme and trehalose 
in the anhydrous LTG mixture at w$_{\textrm{G}}$ = 5 \%, which may explain why the MSD of lysozyme at 300 K is minimal in this
system (Figure~\ref{msd}a-b), 
since glycerol locates primarily at the protein-sugar interface and is therefore likely to improve the protein-matrix coupling. 
Finally, we analyzed glycerol molecules that bridge the two rings of trehalose (see Figures S12 and S13 in the Supporting Information), 
following the hypothesis made by Dirama~et al.~\cite{Dirama2005} that such bridges would constrain trehalose motions.
However, we did not find any straightforward relationship between such HB patterns and the effect of glycerol on the fast dynamics 
of trehalose.

\subsubsection{Geometry of HBs}

\begin{figure}[htbp]
\includegraphics[width=7.5cm,clip=true]{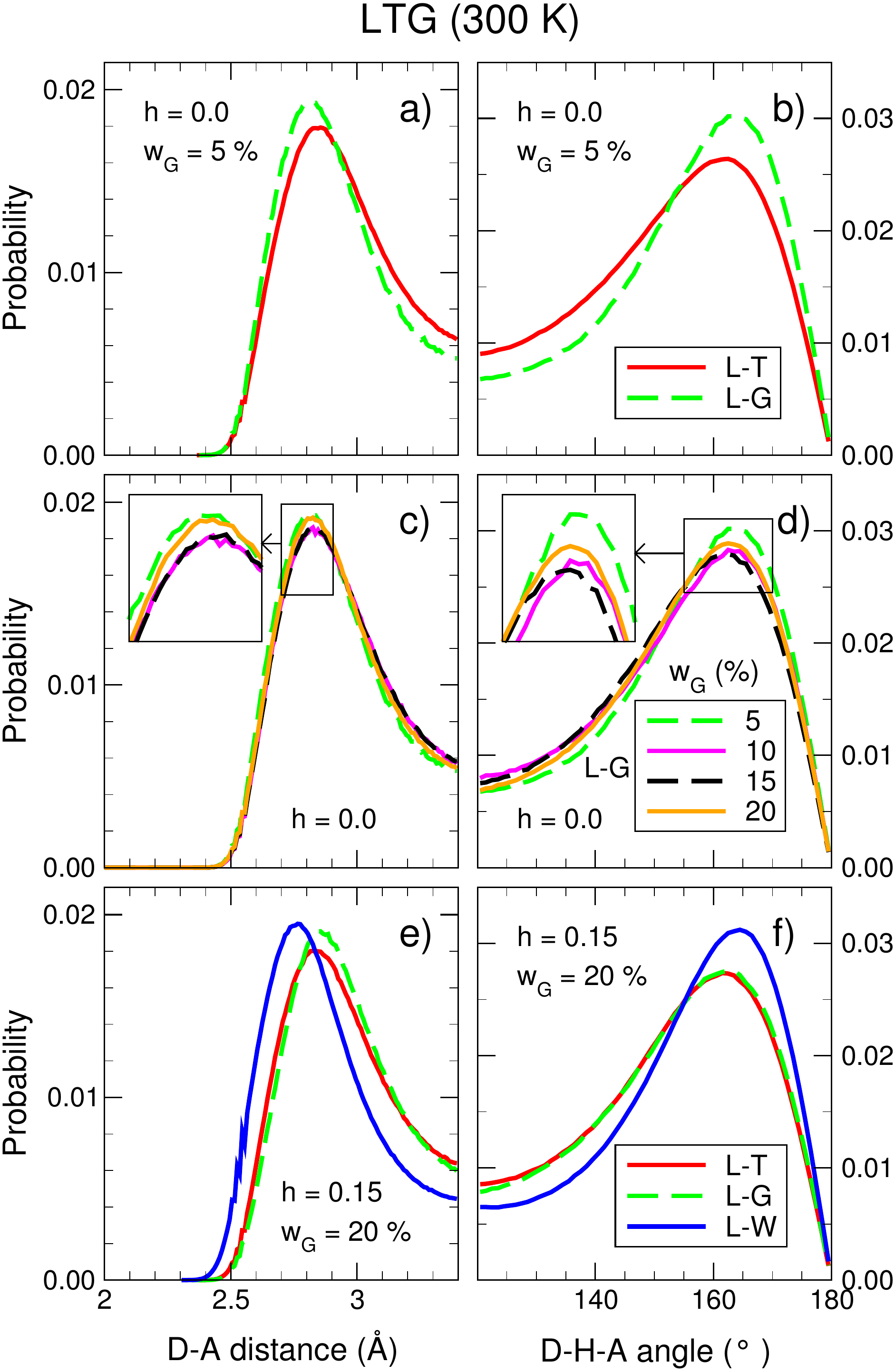}
\caption{\label{hb_dist_ang}
Distributions of the donor-acceptor distance (left panel) and donor-hydrogen-acceptor angle (right panel) of various HBs
formed in the LTG matrices at 300~K: (a-b) L-T and L-G HBs at h = 0.0 and w$_{\textrm{G}}$ = 5 \%, (c-d) L-G HBs at h = 0.0 and 
w$_{\textrm{G}}$ = 5, 10, 15, and 20 \%, (e-f) L-T, L-G, and L-W HBs at h = 0.15 and w$_{\textrm{G}}$ = 20 \%.
Rather similar distributions were observed at 100 K (see Figure S14 in the Supporting Information).
}
\end{figure}

The chemical heterogeneity of the polar groups involved in HBs (hydroxyls, carbonyls, carboxylates, etc.) implies that the HBs
formed by lysozyme, trehalose, glycerol, and water are characterized by various geometries, strengths, and dynamics that
are not accounted for by the analysis performed above. In Figure~\ref{hb_dist_ang}, we present the distance and angle 
distributions of various HBs formed in the LTG matrices. 
Interestingly, L-G HBs are, on average, shorter and more linear than L-T ones in the 
anhydrous LTG matrix at w$_{\textrm{G}}$ = 5 \% (Figure~\ref{hb_dist_ang}a-b). Given that the hydroxyl hydrogen and oxygen 
atoms of glycerol and trehalose share the very same non-bonded parameters in the 
CHARMM36 force field used in the present study~\cite{Guvench2008,Guvench2009,Hatcher2009}, 
these differences in the geometry of HBs reflect differences in sterical and/or 
topological constraints. This supports that glycerol interacts more intimately with lysozyme than trehalose does, 
owing to its smaller size and less complex topology. This result probably explains why the addition of glycerol reduces the MSD of 
lysozyme, particularly at the ps time scale (see Figure~\ref{msd}a-b). Figure~\ref{hb_dist_ang}c-d reveals that the distance and angle
distributions of L-G HBs in LTG matrices at $h$ = 0.0 hardly change with the glycerol content. Nonetheless, it is
interesting to notice that the L-G HBs are very sligthly shorter and/or more linear, and therefore slightly stronger, 
for w$_{\textrm{G}}$ = 5 \%. This may actually rationalize the minimum of MSD(L) at w$_{\textrm{G}}$ = 5 \% found in
Figure~\ref{msd}a. Finally, Figure~\ref{hb_dist_ang}e-f evidences that L-W HBs are significantly shorter and more linear, and thus much stronger, than 
L-T HBs (in line with our previous study~\cite{Lerbret2012}) and than L-G ones. This result confirms the suggestion of Roussenova et 
al.~\cite{Roussenova2014} that water molecules may form stronger HBs than the OH groups of carbohydrates, owing to the fewer 
constraints they experience for their intermolecular interactions. 
It likely explains the \textit{antiplasticizing}
effect of water observed on the fast motions of lysozyme (see Figure~\ref{msd}) and trehalose (we also found that T-W HBs
are stronger than T-T ones, see for instance the results obtained for TG matrices in Figure S15 in the Supporting Information), 
since the addition of water increases both the number 
(see Figure~\ref{nb_HBs} and Table~S4 in the Supporting Information) and the strength of the intermolecular 
HBs formed by lysozyme and trehalose in the LTG matrices.
It must also be pointed out that the stronger L-W HBs compared to L-T and L-G ones 
probably accounts in part for the preferential interaction of lysozyme with water, beyond the straightforward larger 
sterical and topological constraints experienced by trehalose and glycerol compared to water.

\newpage

\subsubsection{Dynamics of HBs}

\begin{figure}[htbp]
\includegraphics[width=6cm,clip=true]{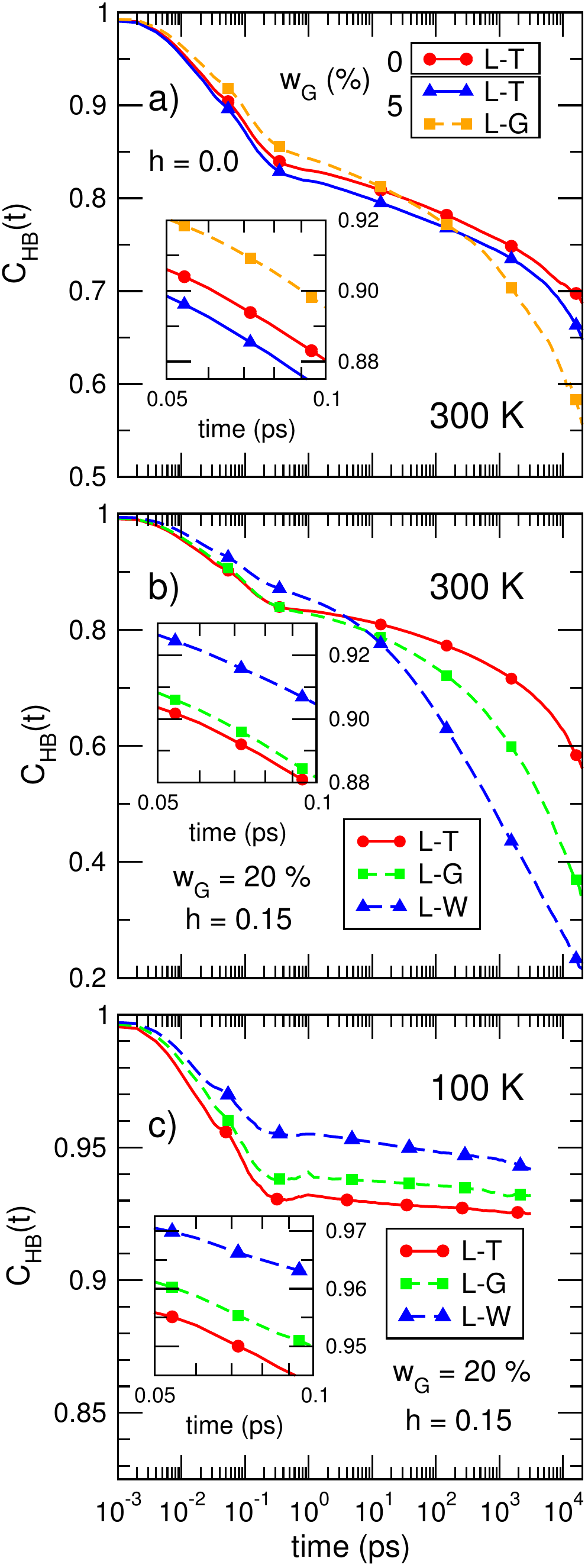}
\caption{\label{hb_dynamics}
Time autocorrelation function, C$_{\textrm{HB}}$, of different types of HBs found in the LTG matrices: 
L-T and L-G HBs at h = 0.0 and w$_{\textrm{G}}$ = 0 or 5 \% at 300 K (a), L-T, L-G, and L-W HBs at h = 0.15 
and w$_{\textrm{G}}$ = 20 \% at 300 K (b) and at 100 K (c). The insets show zooms on the 0.05-0.1 ps 
time range, in which the decay of C$_{\textrm{HB}}$ strongly depends on the strength of the considered HBs.
The C$_{\textrm{HB}}$ of T-T, T-G, and T-W HBs in LTG and TG matrices are provided in Figure S16 
in the Supporting Information.
}
\end{figure}

The various geometries of the HBs formed in LTG and TG matrices imply a high diversity of dynamics, which needs to
be considered to understand the \textit{plasticizing} and \textit{antiplasticizing} effects of glycerol and water on 
the fast motions of lysozyme and trehalose. In this study, we investigated the dynamics of HBs using the time autocorrelation
function, C$_{\textrm{HB}}$, defined as $C_{HB}(t)=<b(0).b(t)>/<b>$, where b(t) is 1 if a 
D-H$\cdot\cdot\cdot$A HB between a given set of donor, D, hydrogen, H, and acceptor, A, atoms exists at time t, and is zero 
otherwise~\cite{Luzar1996}. The brackets mean averaging over the different pairs of HBs and time origins. By definition, 
C$_{\textrm{HB}}$(t) relates to the probability that a HB formed at time 0 still exists at time t, even if it has broken 
in between. The short-time decay of C$_{\textrm{HB}}$(t) originates from fast motions such as librations or intermolecular 
vibrations, and therefore denotes the strength of the HBs considered. In contrast, the long-time behavior of
C$_{\textrm{HB}}$(t) results from the relative diffusion of the donor and acceptor atoms involved in the HBs and, thus, 
describes the structural relaxation of HBs. The slower decay of C$_{\textrm{HB}}$ for L-G HBs than for L-T ones
at time scales shorter than about 10 ps in the anhydrous LTG matrix at w$_{\textrm{G}}$ = 5 \% confirms that L-G HBs 
are stronger than L-T ones in this system (Figure~\ref{hb_dynamics}a), consistent with the results from Figure~\ref{hb_dist_ang}a-b. 
This explains why the addition of small amounts of glycerol decreases MSD(L) at short time scales. Note, however, that 
the faster decay of L-G HBs at longer time scales - and at 300 K - agrees with the plasticizing effect of glycerol usually reported in 
the literature~\cite{Lourdin1998,Verbeek2010}, which is also revealed by the much faster decay of C$_{\textrm{HB}}$ for T-G HBs 
than for T-T ones at long time scales in both LTG and TG matrices at 300 K (see Figure S16 in the Supporting Information).
Moreover, the slowing down of the structural relaxation of T-T, T-G, and T-W HBs induced by the presence of lysozyme 
(Figure S16 in the Supporting Information) suggests that (i) lysozyme confines the motions of trehalose, glycerol, and water 
and, therefore, hinders their diffusion (see \textit{e.g.} the MSDs of glycerol and water in Figure S6 in the Supporting Information), 
and that (ii) the preferential interaction of glycerol and water with lysozyme in comparison with trehalose makes them less likely 
to plasticize trehalose motions. In addition, Figure~\ref{hb_dynamics}b-c confirms that L-W HBs are stronger than L-T and L-G ones, 
in line with the results from Figure~\ref{hb_dist_ang}e-f and with the MD study of Tarek and Tobias on RNase A in glycerol or in 
water~\cite{Tarek2008}, which suggests that protein-water HBs relax more slowly than protein-glycerol ones at sub-ps time scales 
(see Figure 4a of ref.~[\citen{Tarek2008}]). Similarly, T-W HBs appear stronger than T-T and T-G ones in both LTG and TG matrices 
(Figure S16 in the Supporting Information). These results seem in line with a recent thermodynamic analysis of HBs in pure water 
and in a 0.8~m trehalose/water binary solution by Sapir and Harries~\cite{Sapir2017}, who determined that the enthalpic contribution 
to the free energy of HB formation, $\Delta G$, is larger in T-W HBs (-3.6~$\pm$~0.4~kJ.mol$^{\textrm{-1}}$) than in T-T ones 
(-2.8~$\pm$~0.6~kJ.mol$^{\textrm{-1}}$) (note that $\Delta G$ was only very slightly more favorable for T-W HBs than for T-T ones, 
-3.5~$\pm$~0.5~kJ.mol$^{\textrm{-1}}$ vs -3.4~$\pm$~0.9~kJ.mol$^{\textrm{-1}}$ at 298~K). 
We can thus assume that the reduced MSDs of lysozyme and trehalose in the presence 
of water stems from the formation of strong L-W and/or T-W HBs. At 300 K and time scales longer than $\sim$ 1-10 ps, L-T and T-T 
HBs relax much slower than L-G or T-G ones, and even more than L-W and T-W HBs, thereby explaining why the substitution of L-T HBs 
by L-W and L-G ones on one hand and of T-T HBs by T-W and T-G ones on the other hand (see Figure~\ref{nb_HBs}, and Tables~S4 and S5 in the 
Supporting Information) leads to a \textit{plasticizing} effect of glycerol and water on lysozyme and trehalose motions. Conversely, 
L-W (T-W) HBs relax more slowly than L-T (T-T) and L-G (T-G) ones up to the ns time scale at 100 K 
(Figure~\ref{hb_dynamics}c and Figure S16c,f in the Supporting Information), 
owing to the reduced molecular mobility of water at such a low temperature (Figure S6 in the Supporting Information). 
Thus, the more numerous and/or stronger HBs formed by lysozyme and trehalose in 
the presence of water or glycerol rationalizes why the fast dynamics of proteins may be better suppressed in these solvents than in 
trehalose at low temperatures. Indeed, the structural relaxation of protein-water and protein-glycerol HBs at such temperatures 
becomes so slow that the breaking of HBs on the ps-ns time scale then primarily depends on the strength of the HBs formed rather than 
on the diffusion of solvent molecules. Therefore, our results clarify why glycerol and water may exhibit both \textit{plasticizing} 
and \textit{antiplasticizing} effects, depending on the temperature considered, and corroborate previous experimental ones 
from the literature~\cite{Nickels2012,Sakai2013,Caliskan2004}. For example, they are in line with those of Nickels et al., 
which demonstrated that hydration water reduces protein fast motions at low temperatures, but enhances them at higher 
temperatures. The present results also fairly agree with those of Garcia-Sakai et al.~\cite{Sakai2013}, which showed that 
the quasi-elastic intensity of 1:1 L/G and 1:0.5 L/D$_{\textrm{2}}$O samples is weaker than those of dry L and 1:1 L/T 
ones at 150 K. Finally, they are consistent with the neutron and light scattering results of 
Caliskan~et al.~\cite{Caliskan2004}, which evidenced that glycerol is more efficient than trehalose to dampen 
the dynamics of lysozyme at temperatures below $\sim$~270~K, whereas trehalose is more effective at higher temperatures,
as expected from its higher T$_{\textrm{g}}$ ($\sim$~390~K~\cite{Miller2000,Cicerone2004,Weng2015} 
vs $\sim$~190~K for glycerol~\cite{Bohmer1993,Hempel2000}).

\newpage

\section{Conclusions}

This study evidences that glycerol and water at low concentrations modify the structural and dynamical properties 
of LTG and TG matrices to various extents. The addition of small amounts of glycerol and/or water decreases their 
glass transition temperature, T$_{\textrm{g}}$, and improves their molecular packing. 
However, the density of LTG matrices exhibit small maxima at glycerol contents of 5-10 wt.~\%, 
whereas that of TG mixtures steadily decreases when w$_{\textrm{G}}$ increases. This difference may be ascribed 
to the much larger excluded volume effects induced by the presence of lysozyme in LTG matrices, which prevents an 
efficient packing between protein and sugar molecules, given its large size, globular shape, and the roughness of 
its surface. Accordingly, LTG matrices have lower densities, larger free volume fractions and are composed of 
bigger holes than TG ones, and trehalose molecules occupy a larger volume in LTG matrices. Such differences 
modulate the effects of glycerol and water on the fast dynamics of lysozyme and trehalose, which depend on the time 
scale and the temperature considered. At 300 K, glycerol tends to act as an antiplasticizer on lysozyme and 
trehalose motions in LTG matrices by reducing their amplitudes at the ps and the ns time scales, but it mostly
plasticizes trehalose motions in TG matrices. Water antiplasticizes the ps-time scale motions of lysozyme and 
trehalose in LTG and TG matrices (only at $h$ = 0.075 in TG mixtures), and it exhibits a plasticizing effect at 
the ns time scale in both matrices. At 100 K, water systematically antiplasticizes the fast motions of lysozyme 
and trehalose in both LTG and TG systems, whatever the time scale (ps or ns) considered, and glycerol also shows 
such an effect at $h$ = 0.0 and 0.075. The crossovers from a plasticizing effect at room temperature to an 
antiplasticizing one at lower temperatures that we found for water and glycerol at the ns time scale are consistent 
with those reported in previous experimental studies for glycerol in TG mixtures~\cite{Anopchenko2006,Obrzut2010} 
and for glycerol and water in protein matrices~\cite{Nickels2012,Sakai2013}. To understand the origin and the 
temperature dependence of those effects, we characterized the intermolecular HBs formed between species. Lysozyme 
and trehalose form more HBs in presence of glycerol and/or water. Nonetheless, the HBs that lysozyme forms with 
trehalose, glycerol, and water in LTG matrices are much less numerous that the ones it forms with its hydration 
water in dilute aqueous solution. Therefore, the water replacement hypothesis~\cite{Crowe1984b} is only partially 
valid, since trehalose and glycerol cannot perfectly substitute for water upon dehydration. In addition, the 
distribution between species of the intermolecular HBs formed by lysozyme reveals that trehalose, glycerol, and water 
are not distributed homogeneously in the LTG mixtures. Rather, lysozyme interacts preferentially with water, even 
more when the water content decreases, in line with the water anchorage hypothesis, which assumes that residual water
anchors the protein dynamics to its surroundings~\cite{Cordone2005,Francia2008}. Interestingly, our results also 
reveal that the preferential interaction of lysozyme with glycerol rather than with trehalose in the anhydrous LTG 
matrices necessarily stems from the larger size of trehalose and its more complex topology with respect to glycerol, 
since the hydroxyl groups of trehalose and glycerol share the very same non-bonded interaction parameters in our 
simulations. Besides, lysozyme-water HBs are stronger than lysozyme-glycerol and lysozyme-trehalose ones, which 
can be ascribed to the smaller size of water with respect to glycerol and trehalose as well as to its more polar 
O-H bond that allow it to interact more favorably with the protein than glycerol and trehalose do. Moreover, 
lysozyme-water and lysozyme-glycerol HBs relax much faster than lysozyme-trehalose ones at 300 K and at time scales 
greater than $\sim$ 10-100 ps, owing to the larger mobilities of water and glycerol 
compared to that of trehalose, thereby corroborating the plasticizing effects of water and glycerol usually reported 
in the literature. Nonetheless, L-W HBs, and to a lower extent, L-G HBs, relax more slowly than L-T ones
at 100 K, at which the mobilities of water and glycerol are strongly reduced, so that the structural relaxation of
HBs depends to a larger extent on the strength of the HBs formed. This explains why at low temperatures the 
MSD of hydrated proteins may be reduced with respect to that of dry proteins or why the MSD of proteins in trehalose 
glasses may exceed that of proteins in glycerol solvent, as revealed by several experimental 
studies~\cite{Nickels2012,Sakai2013}. Our results thus rationalize why glycerol performs better than trehalose at low 
temperatures~\cite{Caliskan2004,Sakai2013}. Given that the stability of proteins in sugar glasses has been
correlated with the inverse of the MSD of the matrix in neutron scattering experiments~\cite{Cicerone2012}, our results 
clearly suggest that the addition of small amounts of glycerol and water may improve the stability of freeze-dried 
proteins by increasing the number and/or the strength of the HBs they form, thereby stiffening the glassy matrix and reducing protein 
motions that may lead to denaturation.

Overall, the structural properties of the LTG and TG matrices investigated in this work (density, free volume fraction, 
numbers and distributions of HBs, etc.) exhibit dependences on the concentrations of glycerol and water that appear rather
similar at 300~K and at 100~K, in contrast to dynamical properties (MSDs, structural relaxation of HBs). Moreover, 
at a given weight content, water usually influences such properties to a deeper extent than glycerol does, owing to its 
smaller size and more favorable interactions with lysozyme and trehalose. Finally, this study suggests that the potentially 
stabilizing effects of low concentrations of glycerol and water on proteins may probably not be deciphered by considering only 
mixtures of excipients and water, since proteins, excipients, and water mutually influence each other.

%% \suppinfo macro.
%%%%%%%%%%%%%%%%%%%%%%%%%%%%%%%%%%%%%%%%%%%%%%%%%%%%%%%%%%%%%%%%%%%%%

\begin{suppinfo}
Full simulation details; compositions and densities of simulated systems; contents of $\alpha$-helices and $\beta$-sheets of lysozyme; 
procedure for the determination of the glass transition temperature; densities and free volume fractions at 100 K; 
spatial distributions of free volume; discussion on the hole filling mechanism; 
molecular volumes of lysozyme and trehalose; examples of MSDs of lysozyme, trehalose, glycerol, 
and water; correlation between MSD and density; $n_{\textrm{HB}}$(L) and $n_{\textrm{HB}}$(T) in LTG and TG matrices; 
decompositions of $n_{\textrm{HB}}$(L), $n_{\textrm{HB}}$(T), $n_{\textrm{HB}}$(G), and $n_{\textrm{HB}}$(W); 
bridging glycerol and water molecules; geometry and dynamics of HBs in LTG and TG matrices.\\
\end{suppinfo}

%% classes.  Rather than use \section, an appropriate macro is
%% provided that will always work.
%%%%%%%%%%%%%%%%%%%%%%%%%%%%%%%%%%%%%%%%%%%%%%%%%%%%%%%%%%%%%%%%%%%%%
\acknowledgement

Most calculations were performed using HPC resources from PSIUN CCUB (Centre de Calcul de l'Universit\'e de Bourgogne).
Prof. Alain H\'edoux (UMET, Universit\'e de Lille) and Prof. Marc Descamps (UMET, Universit\'e de Lille)
are acknowledged for fruitful discussions.


\bibliography{paper}

\end{singlespacing}

\includepdf[pages=-]{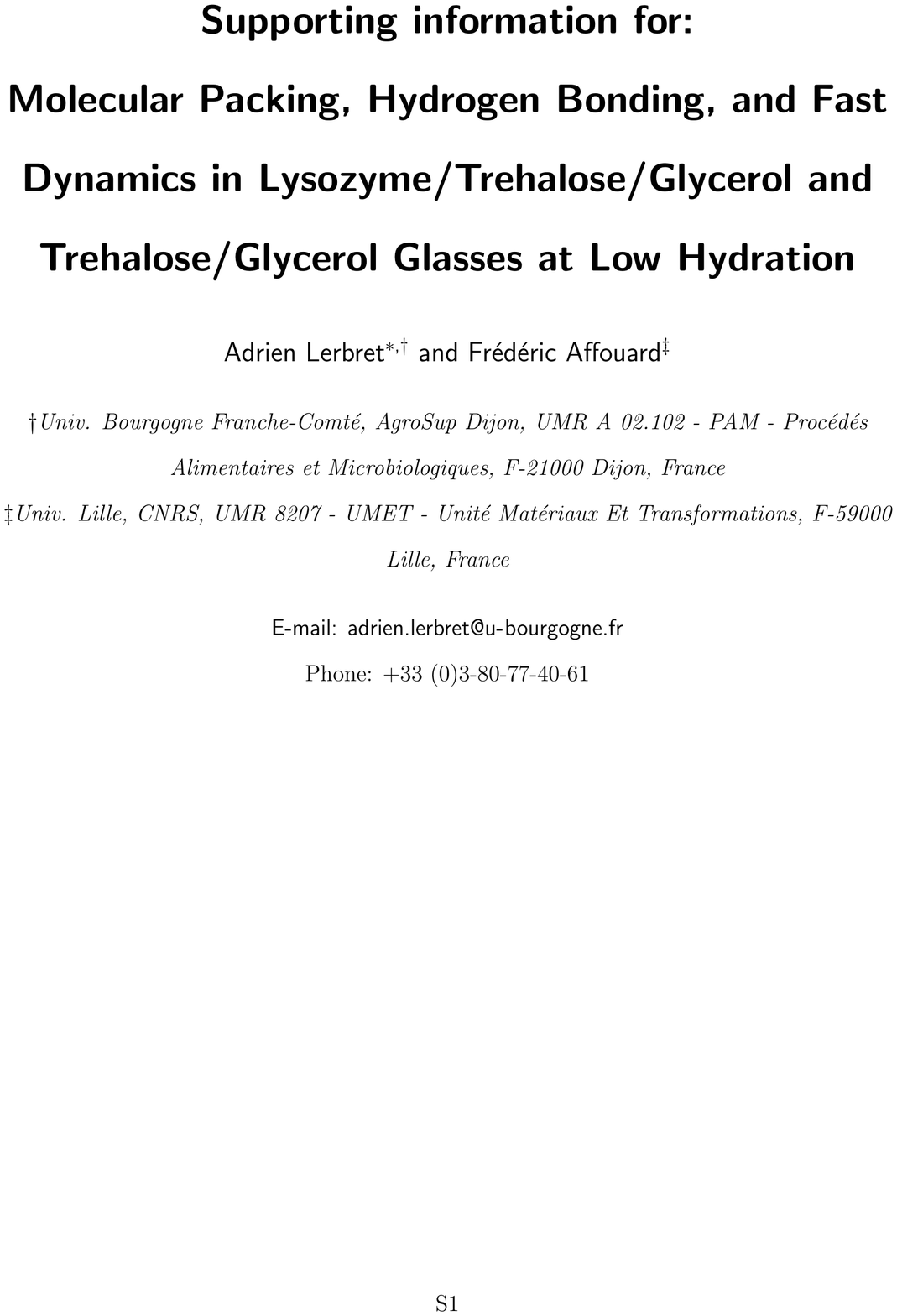}

\includepdf[pages={-},angle=90]{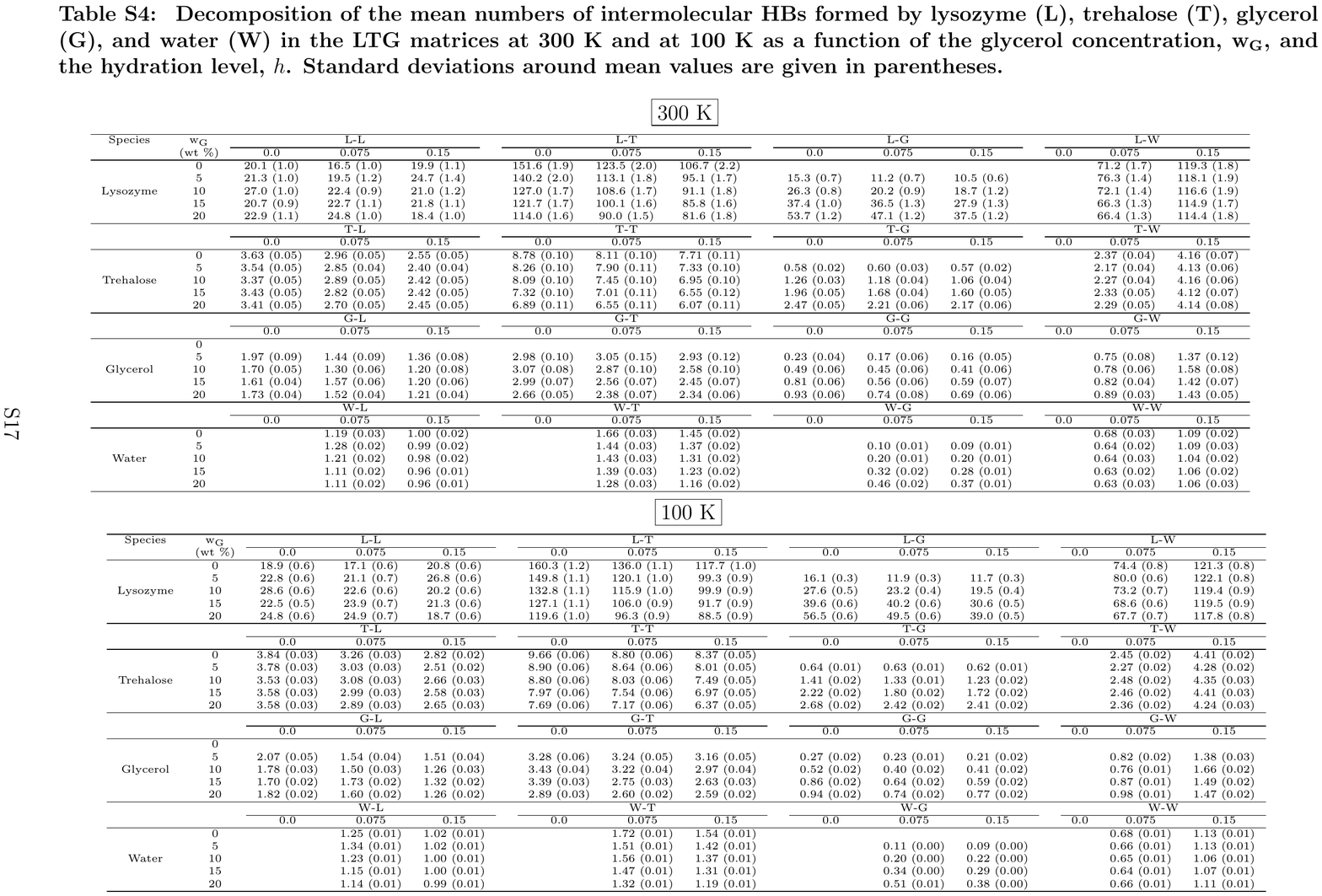}

\includepdf[pages=-]{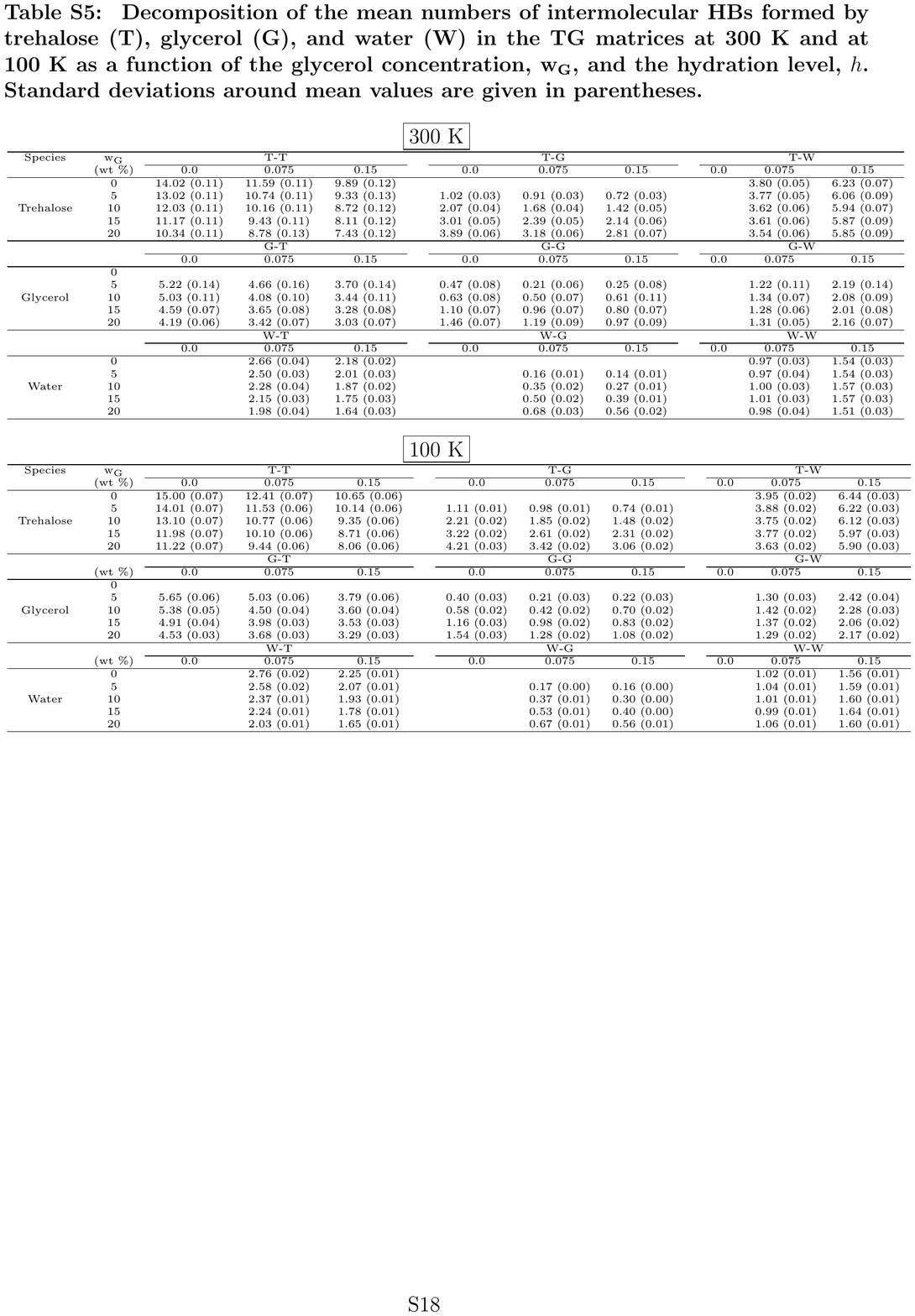}

\end{document}